\documentclass[pra,twocolumn,superscriptaddress,notitlepage]{revtex4-2}

\usepackage{etoolbox}
\newbool{noMarkup}

\booltrue{noMarkup}

\pdfoutput=1
\usepackage{makeidx}
\usepackage[utf8]{inputenc}
\usepackage{graphicx}
\usepackage{amsmath}
\usepackage{amssymb}
\usepackage{mathrsfs}
\usepackage{subfigure}
\usepackage{dcolumn}
\usepackage{bm}
\usepackage{bbm}
\usepackage{color}
\usepackage{esint}
\usepackage{lineno}
\usepackage[breaklinks,
            colorlinks,
            urlcolor=blue,
            linkcolor=blue,
            anchorcolor=blue,
            citecolor=blue]{hyperref}

\usepackage{xcolor}
\definecolor{jfColor}{HTML}{FF00FF}

\definecolor{jbColor}{HTML}{33AAFF}

\usepackage{ulem}
\usepackage{orcidlink}

\ifbool{noMarkup}{
    \usepackage[final]{changes}
    \def\addedStart{}
    \def\addedEnd{}
}{
    \usepackage[markup=underlined]{changes}
    \def\addedStart{\color{blue} }
    \def\addedEnd{\color{black} }
}

\begin{document}

\title{Exploring variational quantum eigensolver ansatzes for the long-range XY model
}

\author{Jia-Bin You\orcidlink{0000-0001-8815-1855}}
\email{you\_jiabin@ihpc.a-star.edu.sg}
\affiliation{Institute of High Performance Computing, A*STAR (Agency for Science, Technology and Research), 1 Fusionopolis Way, \#16-16 Connexis, Singapore 138632}

\author{Dax Enshan Koh\orcidlink{0000-0002-8968-591X}}
\email{dax\_koh@ihpc.a-star.edu.sg}
\affiliation{Institute of High Performance Computing, A*STAR (Agency for Science, Technology and Research), 1 Fusionopolis Way, \#16-16 Connexis, Singapore 138632}

\author{Jian Feng Kong\orcidlink{0000-0001-5980-4140}}
\affiliation{Institute of High Performance Computing, A*STAR (Agency for Science, Technology and Research), 1 Fusionopolis Way, \#16-16 Connexis, Singapore 138632}

\author{Wen-Jun Ding}
\affiliation{Institute of High Performance Computing, A*STAR (Agency for Science, Technology and Research), 1 Fusionopolis Way, \#16-16 Connexis, Singapore 138632}

\author{Ching Eng Png}
\affiliation{Institute of High Performance Computing, A*STAR (Agency for Science, Technology and Research), 1 Fusionopolis Way, \#16-16 Connexis, Singapore 138632}

\author{Lin Wu}
\affiliation{Institute of High Performance Computing, A*STAR (Agency for Science, Technology and Research), 1 Fusionopolis Way, \#16-16 Connexis, Singapore 138632}
\affiliation{Science, Mathematics and Technology (SMT), Singapore University of Technology and Design (SUTD), 8 Somapah Road, Singapore 487372}

\begin{abstract}

    Finding the ground state energy and wavefunction of a quantum many-body system is a key problem in quantum physics and chemistry. We study this problem for the long-range XY model by using the variational quantum eigensolver (VQE) algorithm. We consider VQE ansatzes with full and linear entanglement structures consisting of different building gates: the CNOT gate, the controlled-rotation (CRX) gate, and the two-qubit rotation (TQR) gate. We find that the full-entanglement CRX and TQR ansatzes can sufficiently describe the ground state energy of the long-range XY model. In contrast, only the full-entanglement TQR ansatz can represent the ground state wavefunction with a fidelity close to one. In addition, we find that instead of using full-entanglement ansatzes, restricted-entanglement ansatzes where entangling gates are applied only between qubits that are a fixed distance from each other already suffice to give acceptable solutions. Using the entanglement entropy to characterize the expressive powers of the VQE ansatzes, we show that the full-entanglement TQR ansatz has the highest expressive power among them.

\end{abstract}

\maketitle

\section{Introduction}

Simulating quantum many-body systems composed of many particles and learning their properties are, in general, hard problems for classical computers. In contrast, many of these problems may be solved more efficiently by appropriately designed quantum algorithms \cite{doi:10.1021/acs.chemrev.8b00803,PhysRevLett.83.5162,Berry2018,PhysRevA.79.032316,Jordan1130,Temme2011} running on large-scale fault-tolerant quantum computers, which promise to efficiently simulate these systems and provide important insights into their properties \cite{Buluta108,RevModPhys.86.153,Brown2010}. However, such fault-tolerant quantum computers currently still do not exist. The most powerful quantum computers we have today are noisy intermediate-scale quantum (NISQ) \cite{Preskill2018quantumcomputingin,Arute2019} devices that are too small and noisy to run many quantum algorithms, including quantum phase estimation \cite{PhysRevLett.83.5162,Aspuru-Guzik1704}, which require fully coherent evolution. Nevertheless, these devices have already succeeded at solving certain sampling problems believed to be intractable for classical computers \cite{arute2019quantum,zhong2020quantum,wu2021strong}. To harness the full power of these devices to solve a larger number and range of problems, it is essential to ensure efficient management of a limited number of qubits with finite coherence times.

% VQE algorithm in quantum computing NISQ literature survey
%% theory
Hybrid classical-quantum algorithms---such as variational quantum algorithms (VQAs) \cite{Nat.Commun.5.4213,PhysRevA.92.042303,PhysRevResearch.1.023025,Higgott2019variationalquantum,Kokail2019,PhysRevA.99.062304,McArdle2019,PhysRevX.7.021050,sa2021towards,PhysRevX.8.011021,Romero_2017,Carolan2020,Arrasmith2019,LaRose2019,Khatri2019quantumassisted,PhysRevA.101.062310,BravoPrieto2020scalingof,Cirstoiu2020,Sharma2020,PerezSalinas2020datareuploading,cerezo2021variational,PhysRevLett.122.140504,koh2022foundations,PRXQuantum.2.010346,wiersema2021exploring,kattemolle2021variational,bosse2022probing,consiglio2021variational}---leverage classical resources to reduce the required number of quantum gates, and hold the promise of outperforming current purely classical techniques. \addedStart{}The structure of the VQA ansatzes generally depend on the tasks at hand as well as the quantun resources available \cite{cerezo2021variational}. For example, the hardware-efficient ansatz \cite{Nature.549.242} aims at reducing the circuit depth; the unitary coupled ansatz \cite{Nat.Commun.5.4213} is used for problems in quantum chemistry; and the quantum approximate optimization algorithm (QAOA) ansatz is used to obtain approximate solutions to combinatorial optimization problems \cite{Farhi2014, hadfield2019from}.\addedEnd{} The central idea of VQAs is to design a parameterized quantum circuit that depends on a set of gate parameters and whose architecture is dictated by the structure of the NISQ devices \cite{BravoPrieto2020scalingof}. These variational parameters can be optimized using quantum-classical optimization loops by extremizing a cost function which is designed to encode the solution of the optimization problem into its extremum. \addedStart{}In theory, the approximation improves with increasing ansatz depth, though a larger number of noisy gates and variational parameters undermine performance in practice \cite{Herrman2022}. Noise grows rapidly with circuit depth and affects the fidelity of the prepared quantum state, and so shallow circuits could potentially achieve a better performance on current NISQ devices compared to their deeper counterparts.
\addedEnd{}

Here we focus on the variational quantum eigensolver (VQE) \cite{Nat.Commun.5.4213,PhysRevA.92.042303,PhysRevResearch.1.023025,Higgott2019variationalquantum,Kokail2019,PhysRevA.99.062304,McArdle2019,PhysRevX.7.021050,sa2021towards}, a VQA that is designed to approximate the ground state of quantum many-body systems. The VQE---one of the most promising near-term applications using NISQ devices---works by preparing a quantum many-body state on a parameterized quantum circuit and executing the circuit without quantum error correction, which works if the circuit is short enough and the fidelity of gates high enough \cite{PhysRevA.92.042303}.
The parameterized quantum circuit provides a variational ansatz for the ground state, and a classical optimizer tunes the circuit parameters to give a quantum circuit that rotates the product state $\otimes_{n}|0_{n}\rangle$ into (an approximation of) the ground state of a given Hamiltonian of an $n$-qubit quantum many-body system \cite{PhysRevResearch.1.023025}. The efficiency of optimization depends on the number of iterations required for convergence and the number of gates involved in each preparation and measurement cycle of the quantum subroutine \cite{Romero_2018}.
%% experiment
To date, several small-scale experimental demonstrations of the VQE for molecules and quantum magnets have been performed \cite{PhysRevX.6.031007,PhysRevX.8.011021,PhysRevX.8.031022,PhysRevA.95.020501,Nature.549.242}
on various quantum platforms, including photonic chips \cite{Nat.Commun.5.4213}, ion traps \cite{PhysRevA.95.020501} and superconducting circuits \cite{PhysRevX.6.031007,Nature.549.242}.

In this work, we apply the VQE algorithm to search for the ground state energy and wavefunction of the long-range XY model with length $N$. The Hamiltonian of the system can be written as 
\begin{align}
\label{eq:xy_hamiltonian}
H=-J\sum_{i<j}(X_{i}X_{j}+Y_{i}Y_{j})-h\sum_{i}Z_{i},
\end{align}
where $J$ is the coupling strength; $h$ is the Zeeman field; and $X_{i}$, $Y_{i}$, $Z_{i}$ denote the Pauli matrices at site $i$. Note that this model can be used to describe Bardeen-Cooper-Schrieffer (BCS) superconductivity \cite{PhysRevLett.96.230403,PhysRev.112.1900}, which is an important prototype for modern condensed matter physics.

\addedStart{}In general, the ground state of a spin chain can be expressed as a collection of Bloch vectors defined by $(\langle{X_{i}}\rangle,\langle{Y_{i}}\rangle,\langle{Z_{i}}\rangle)$ on the Bloch sphere for each site, as well as the correlations between sites, such as $\langle{X_{i}X_{j}}\rangle$, $\langle{Y_{i}Z_{j}}\rangle$ and so on. The Bloch vector at site $i$ can be rotated from initial state $|0_{i}\rangle$ by single-qubit rotation gates and the correlations in the ground state can be generated by the entangler which consists of two-qubit gates. To fully describe the ground state of the spin chain, a mean field layer and an entangler should be included in an ansatz. 
\addedEnd{}

We first analytically obtain a mean-field solution for the ground state energy based on the mean-field ansatz. The mean-field solution, which neglects entanglement, will serve as a benchmark for other VQE ansatzes that take entanglement into account. We then introduce quantum entanglement
into the VQE ansatz by constructing entanglers with different building gates and different entanglement structures. More specifically, we consider the following three building gates: (i) CNOT gate, (ii) parametric controlled-rotation (CRX) gates, and (iii) parametric two-qubit rotation (TQR) gates. \addedStart{}Note that we gradually increase the number of variational parameters for the three ansatzes, allowing for finer-grained control over the minimization of the Hamiltonian.\addedEnd{} We consider the following two entanglement structures for each of these building gates: (a) linear entanglement and (b) full entanglement. 

Amongst the above ansatzes, we find that the only ansatzes that well describe the ground state energy and wavefunction of the long-range XY model are the full-entanglement CRX and TQR ansatzes; for these two ansatzes, linear entanglement is insufficient. However, we proceed to show that full entanglement is, in fact, not necessary to describe well the ground state energy and wavefunction of the Hamiltonian: it suffices to use CRX and TQR ansatzes in which the entangling gates are restricted to act only between qubits that are a fixed distance away from each other. Using these restricted-entanglement ansatzes in lieu of their full-entanglement counterparts reduces the overall running time of the optimization step in the VQE algorithm. \addedStart{}In addition, we find that the TQR ansatz can reduce the circuit depth by increasing the number of variational parameters introduced in each layer. Increasing the number of parameters allows for finer-grained control over the minimization of the Hamiltonian. While introducing more parameters within one layer, a corresponding reduction in circuit depth preserves the overall performance of the minimization problems.\addedEnd{} Finally, to elucidate the differences in performance of the different ansatzes, we use the entanglement entropy to characterize their expressive powers. We find that the full-entanglement TQR ansatz, which gives a good VQE performance, also has the largest expressive power amongst all the ansatzes considered.

\section{VQE algorithm}

\begin{figure*}[tbp]
\includegraphics[width=18cm]{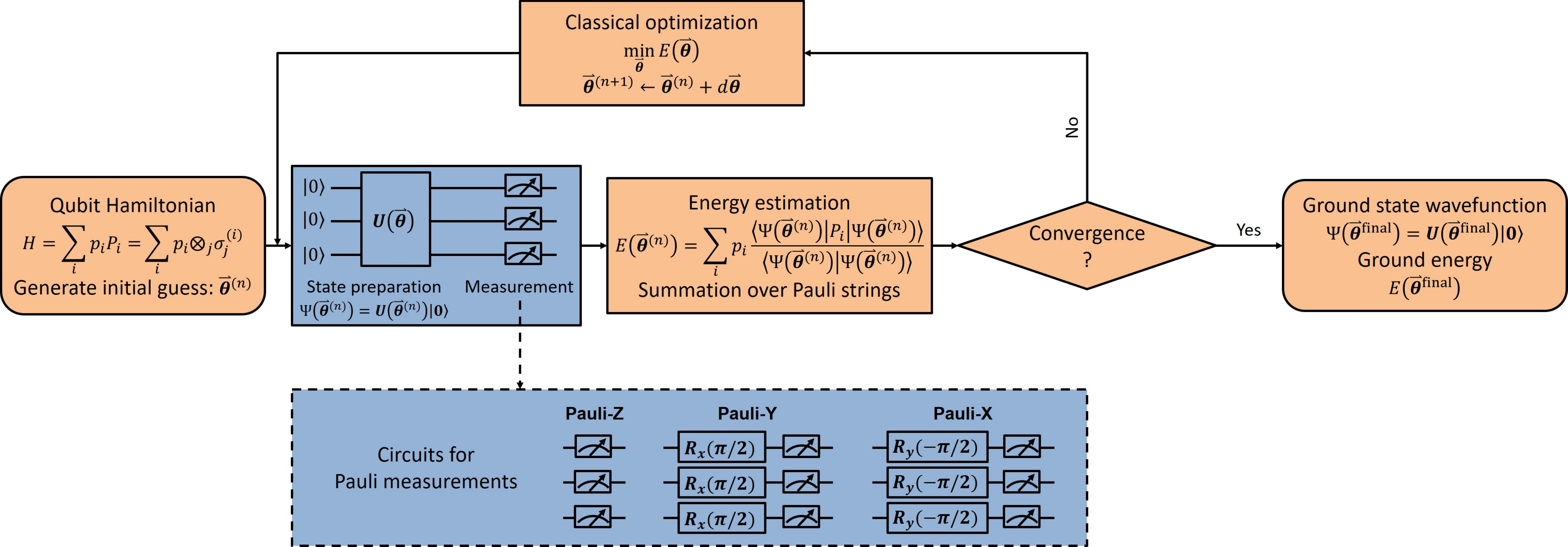}
\caption{Schematic of the VQE algorithm. The long-range XY model is recast into a summation of Pauli strings. An initial guess for the variational parameters is input into the quantum circuit. The estimation of variational energy is obtained by a series of Pauli measurements from the circuits. The classical optimization routine estimates a new value for the circuit parameters. The process is repeated until achieving convergence on the total energy. The red blocks summarize the operations performed by the classical computer, whereas the blue blocks summarize the operations performed by the quantum computer.}
\label{fig1}
\end{figure*}

A schematic of the VQE algorithm is shown in Fig.~\ref{fig1}. A general spin model Hamiltonian $H$ can be represented as a sum of Pauli strings, viz.~$H=\sum_{i}p_{i}P_{i}$, where the $i$-th Pauli string in the sum is $P_{i}=\otimes_{j}\sigma_{j}^{(i)}$, where $\sigma_{j}$ denotes a Pauli matrix (i.e., either $\mathbbm{1}$, $X$, $Y$, or $Z$) acting on site $j$. The ground state of $H$ can be approximated by a quantum circuit ansatz realized by applying a parameterized unitary operation $U(\vec{\theta})$---built from a series of parameterized single- and two-qubit gates---to the computational basis state $\otimes_{j}|0_{j}\rangle$, where $\vec{\theta}$ are the variational parameters of the circuit. We denote the ansatz from the above construction by $|\Psi(\vec{\theta})\rangle=U(\vec{\theta})\otimes_{j}\!|0_{j}\rangle$. 
A guess for the variational parameters $\vec{\theta}^{(0)}$ is inputted into the circuit for initialization. The variational energy $E(\vec{\theta})$ of $H$ can be estimated by performing a series of Pauli measurements to obtain the expectation values $\langle\Psi(\vec{\theta})|P_{i}|\Psi(\vec{\theta})\rangle$, which are then plugged into the formula $E(\vec{\theta})=\sum_{i}p_{i}\langle\Psi(\vec{\theta})|P_{i}|\Psi(\vec{\theta})\rangle/\langle\Psi(\vec{\theta})|\Psi(\vec{\theta})\rangle$. An optimization subroutine is performed to minimize the variational energy (i.e.~we solve the optimization problems $\min_{\vec{\theta}}[E(\vec{\theta})]$ and $\mathrm{argmin}_{\vec{\theta}}[E(\vec{\theta})]$) to find the ground state energy and wavefunction. This subroutine is performed by a classical optimizer, which updates new values for the circuit parameters in each step, via $\vec{\theta}^{(n+1)}\leftarrow\vec{\theta}^{(n)}+d\vec{\theta}$.
This process is repeated until convergence of the ground state energy is achieved. The ground state energy and wavefunction can be calculated as $E(\vec{\theta}^{\text{final}})$ and $|\Psi(\vec{\theta}^{\text{final}})\rangle=U(\vec{\theta}^{\text{final}})\otimes_{j}|0_{j}\rangle$. It should be noted that the classical optimizer may get stuck in a local minimum and fail to find the global minimum for the VQE ansatz. To address this issue, we run the VQE algorithm repeatedly starting from different random parameter seeds. While this approach does not guarantee that the global minimum is found, from a practical point of view it increases the odds of finding an acceptable solution, at the expense of potentially having to run the algorithm an increased number of times. In Fig.~\ref{fig1}, the red blocks indicate the operations performed by the classical computer, whereas the blue blocks indicate the operations performed by the quantum computer.

Pauli measurements are performed on the ansatz $|\Psi(\vec{\theta})\rangle$ to read out the variational energy during the energy estimation subroutine. The expectation of a Pauli string $P_{i}$, $\langle{P_{i}}\rangle=\langle\Psi(\vec{\theta})|P_{i}|\Psi(\vec{\theta})\rangle/\langle\Psi(\vec{\theta})|\Psi(\vec{\theta})\rangle$ can be reconstructed from the countings of computational basis states $\{|0_{j}\rangle,|1_{j}\rangle\}$ of $|\Psi(\vec{\theta})\rangle$ by Pauli measurements $\{\langle{X_{j}}\rangle,\langle{Y_{j}}\rangle,\langle{Z_{j}}\rangle\}$ in the $x$, $y$ and $z$ directions. Our simulations were implemented using \texttt{Qiskit}, IBM's open-source framework for quantum computing \cite{qiskit}.
%In this work, we obtain our results by using \texttt{Qiskit} from IBM. 
In particular, we used the \texttt{QASM Simulator} backend in Aer for all our simulations, except for Fig.~\ref{fig6} and the fidelity calculations, where we used the \texttt{Statevector Simulator} backend in Aer.
The Pauli-Z measurement can simply be obtained by counting the number of states $|0_{j}\rangle$ and $|1_{j}\rangle$ at qubit $j$ and calculating the difference, $\langle{Z_{j}}\rangle=(N_{|0_{j}\rangle}-N_{|1_{j}\rangle})/N_{s}$, where $N_{s}$ is the number of shots in simulations.
The Pauli-X and Pauli-Y measurements can be achieved by first rotating $x$ and $y$ axes to $z$ axis, equivalent to applying $R_{y}(-\pi/2)$ and $R_{x}(\pi/2)$ gates to the circuits respectively, then subsequently performing the Pauli-Z measurement. Here the single-qubit rotation gates are defined as $R_{y}(\theta_{j})=e^{-i\theta_{j}Y_{j}/2}$ and $R_{x}(\theta_{j})=e^{-i\theta_{j}X_{j}/2}$ respectively. The measurement circuits are shown in Fig.~\ref{fig1}. Similarly, we can obtain the expectations of the products of two Pauli operators, such as $\{X_{i}X_{j},Y_{i}Y_{j},Z_{i}Z_{j},X_{i}Y_{j},Y_{i}Z_{j},Z_{i}X_{j}\}$, by rotating the $x$ and $y$ axes to $z$ axis, resulting in the calculation of the expectation of $\langle{Z_{i}Z_{j}}\rangle$ which can be achieved by $\langle{Z_{i}Z_{j}}\rangle=(N_{|0_{i}0_{j}\rangle}-N_{|0_{i}1_{j}\rangle}-N_{|1_{i}0_{j}\rangle}+N_{|1_{i}1_{j}\rangle})/N_{s}$.

\section{Analytical solution for the long-range XY model}

Before discussing the VQE algorithm, we will first give an analytical solution for the isotropic long-range XY model. By using the collective Pauli operators, $S_{x}=\sum_{i}X_{i}$, $S_{y}=\sum_{i}Y_{i}$, and $S_{z}=\sum_{i}Z_{i}$, the Hamiltonian can be recast as
\begin{equation}
\begin{split}
H=-2JS^{+}S^{-}+(J-h)S_{z}+JN,
\end{split}    
\end{equation} where $S^\pm=\frac{1}{2}(S_{x}\pm i S_{y})$. Note that $[S^{+}S^{-},S_{z}]=0$, and so the Hamiltonian and the particle number operator $S^{+}S^{-}$ have the same eigenstates. We can thus represent the wavefunctions in different excitation subspaces. It is found that in the $n$-excitation subspace, the symmetric state, $\psi_{\text{sym}}\sim\sum_{\{i_{1},i_{2},\ldots,i_{n}\}}\sigma_{i_{1}}^{+}\sigma_{i_{2}}^{+}\cdots\sigma_{i_{n}}^{+}|\mathbf{0}\rangle$, has the lowest energy per site,
\begin{equation}
\begin{split}
E_{\text{sym}}^{(n)}/N=(1-2n/N)h-2Jn(1-n/N).
\end{split}
\end{equation}
Here $\sigma_{i_{k}}^{+}$ is the raising operator at site $i_{k}$ and $\{i_{1},i_{2},\ldots,i_{n}\}$ are all the possible combinatorics from $\{0,1,\ldots,N-1\}, (0\le{n}\le{N})$. Thus the ground state energy per site of the long-range XY model is $E_{\text{GS}}/N=\min\{E_{\text{sym}}^{(n)}/N,(n=0,1,\ldots,N)\}$.

The quantum phase transitions occur at the critical points defined by $E_{\text{sym}}^{(n+1)}/N=E_{\text{sym}}^{(n)}/N$. After some calculations, we find that the critical point for transition from $n$-excitation phase to $n+1$-excitation phase is $h/J=2n+1-N$, where $n$ ranges from $0$ to $N-1$. The various transitions are essentially of the same nature, i.e., transition from $n$ excitations to $n\pm1$ excitations, equivalently, $\langle{S_{z}}\rangle=s$ to $s\pm 2$ sectors. $\langle{S_{z}}\rangle$ changes by 2 every time across each transition, and thus the slope $dE_{\text{GS}}/dh$ changes by 2 each time. When $N$ is odd, the ground state magnetization cannot be zero and is always in the paramagnetic phases with different polarizations. Interestingly, when $N$ is even, there exists a ferromagnetic phase when $|{h/J}|\le1$ and the net magnetization is zero. The system undergoes a ferromagnetic to paramagnetic phase transition across the critical point $|h/J|=1$. We will focus on this phase transition behavior in the rest of the paper.

\section{VQE algorithm for mean field ansatz}

\begin{figure}[tbp]
\includegraphics[width=5.5cm]{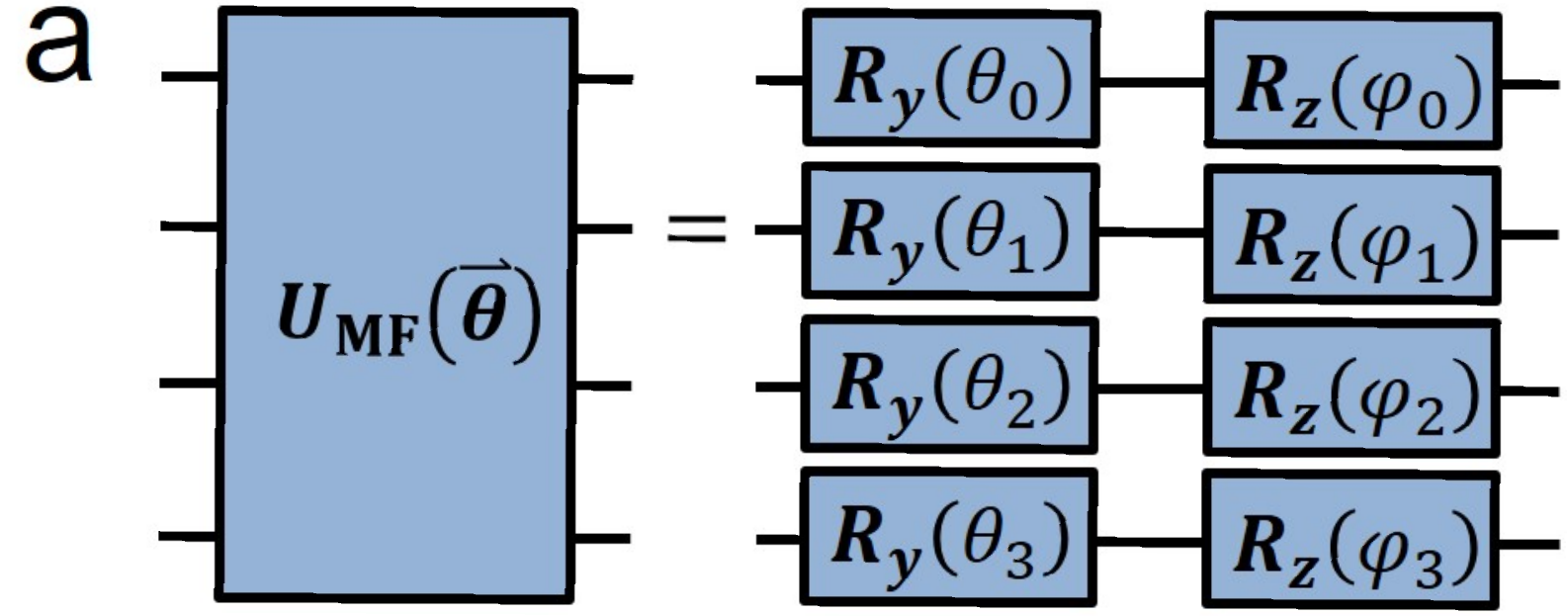}\\
\includegraphics[width=5.5cm]{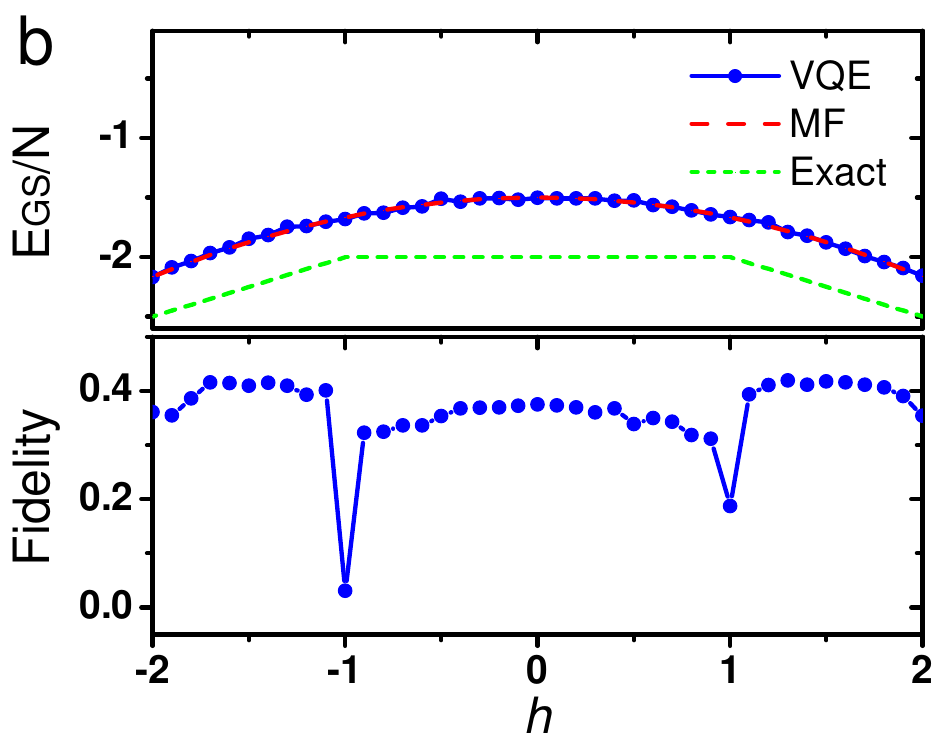}\\
\caption{VQE algorithm with mean-field ansatz. a: quantum circuits for the mean-field ansatz. b: ground state energy per site $E_{\text{GS}}/N$ and fidelity between VQE algorithm and exact solution as a function of the Zeeman field $h$. The parameters are $J=1$, $N=4$. The number of shots is $2^{14}$ in all simulations.}
\label{fig2}
\end{figure}

We first consider the mean-field ansatz where the ground state is approximated as a product state of single qubits,
\begin{equation}
\label{MF}
\begin{split}
|\Psi_{\text{MF}}\rangle=\bigotimes_{p}\left[\begin{array}{*{20}ccccccccc}
{\cos{\tfrac{\theta_{p}}{2}}}\\
{\sin{\tfrac{\theta_{p}}{2}}e^{i\varphi_{p}}}\\
\end{array}\right].
\end{split}
\end{equation}
Notice that there are no correlations within the mean-field ansatz. We can thus decouple the hopping terms and write the mean-field variational energy as $E_{\text{MF}}=-J\sum_{p<q}(\langle{X_{p}}\rangle\langle{X_{q}}\rangle+\langle{Y_{p}}\rangle\langle{Y_{q}}\rangle)-h\sum_{p}\langle{Z_{p}}\rangle$. In the mean-field ansatz, the expectations of Pauli matrices are straightforwardly calculated as $\langle{X_{p}}\rangle=\sin{\theta_{p}}\cos{\varphi_{p}}$, $\langle{Y_{p}}\rangle=\sin{\theta_{p}}\sin{\varphi_{p}}$, and $\langle{Z_{p}}\rangle=\cos{\theta_{p}}$, which can be represented as a point on the Bloch sphere. Therefore, we can obtain an analytical solution for the mean-field variational energy,
\begin{equation}
\label{MFenergy}
\begin{split}
E_{\text{MF}}&=-J\sum_{p<q}\sin{\theta_{p}}\sin{\theta_{q}}\cos{(\varphi_{p}-\varphi_{q})}-h\sum_{p}\cos{\theta_{p}}.\\
\end{split}
\end{equation}
Minimizing the energy $E_{\text{MF}}$ with respect to the variational parameters $\{\theta_{p},\varphi_{p}\}$, we can estimate the ground state energy within the mean-field variational class.

On the other hand, the mean-field ansatz in Eq.~\eqref{MF} can readily be implemented by the quantum circuit shown in Fig.~\ref{fig2}a. This state can be realized by sequentially applying two single-qubit rotation gates $R_{y}(\theta_{p})$ and $R_{z}(\varphi_{p})$ on each qubit $p$ initially prepared in the state $|0_{p}\rangle$. We can then use VQE to estimate the ground state energy. We first perform the Pauli measurements to evaluate the variational energy $E_{\text{MF}}$, and then update the variational parameters $\{\theta_{p},\varphi_{p}\}$ to minimize the variational energy $E_{\text{MF}}$ by a classical optimizer and finally to obtain the ground state energy. Notice that for the $N$-qubit mean-field ansatz of the long-range XY model, we have $N^2$ Pauli strings (corresponding to the number of terms in the Hamiltonian) and $3N$ Pauli measurements $\{\langle{X_{p}}\rangle,\langle{Y_{p}}\rangle,\langle{Z_{p}}\rangle\}$ $(p=0,1,\ldots,N-1)$ to evaluate the expectation of Hamiltonian.

The results for the mean field ansatz are shown in Fig.~\ref{fig2}b. In the upper plane of Fig.~\ref{fig2}b, we study the ground state energy per site $E_{\text{GS}}/N$ as a function of Zeeman field $h$. The blue-solid, red-dashed, and green-dotted lines result from the VQE algorithm, the mean-field solution and the exact solution respectively. It is found that the VQE solution based on the mean-field ansatz is consistent with the mean-field solution. However, due to the ignorance of intrinsic correlations in the long-range XY model, there is a deviation between the VQE solution and the exact solution, which is also evidenced by the low fidelity $F=|\langle\psi_{\text{VQE}}|\psi_{\text{exact}}\rangle|^2<0.5$ between the VQE algorithm and the exact solution shown in the lower plane of Fig.~\ref{fig2}b. The fidelity becomes even worse near the critical point of phase transition $h=1$. The asymmetry of the energy and the fidelity results from the sampling error of the \texttt{qasm} simulator in IBM \texttt{qiskit} package. This sampling error will induce the error in estimating the gradient descent in the classical optimization routine which is liable to become trapped in local minima.

\section{VQE algorithm with CNOT ansatz}
\label{sec:VQE_algorithm_with_CNOT_ansatz}

\begin{figure}[tbp]
\includegraphics[width=6cm]{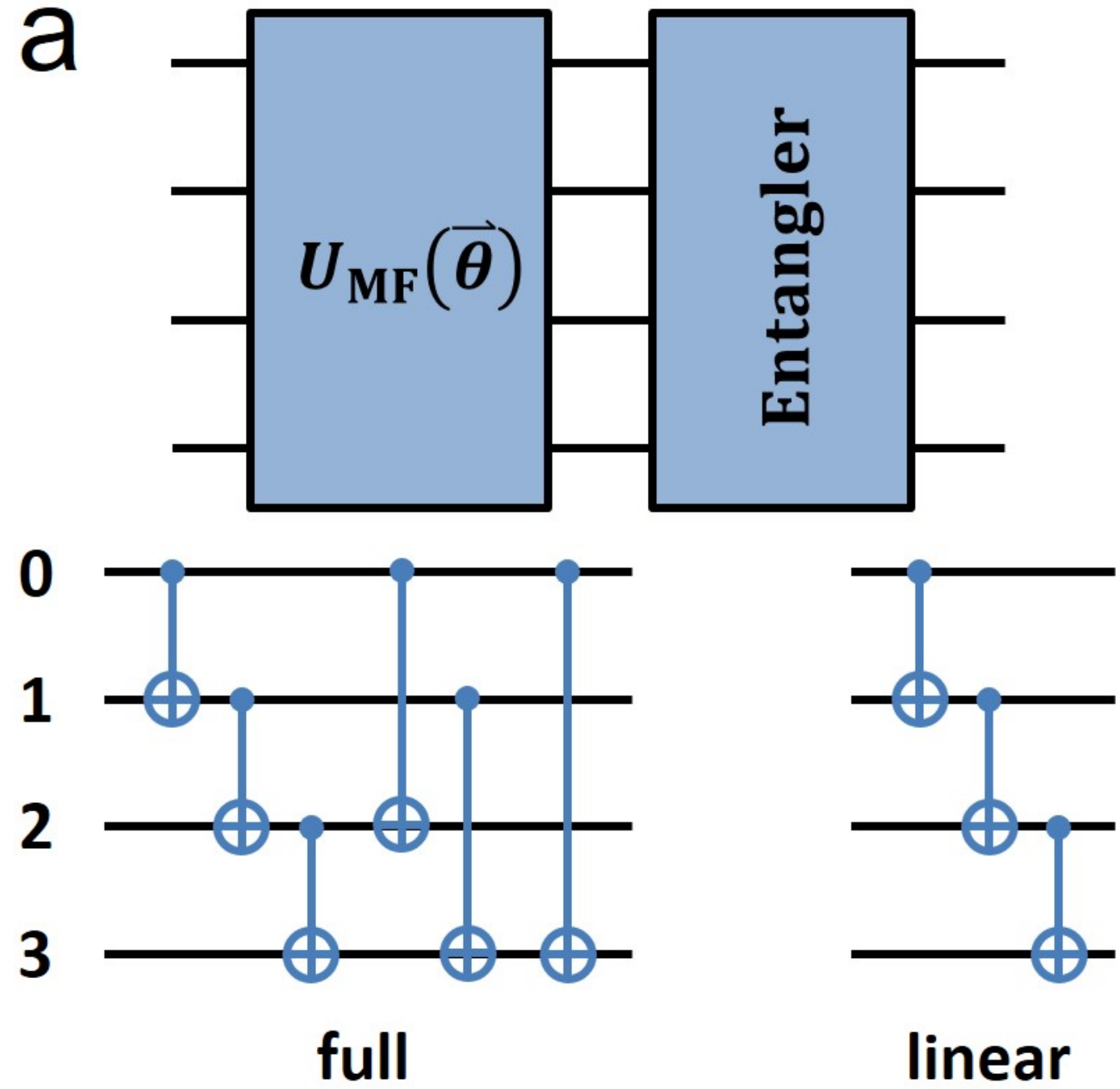}\\
\includegraphics[width=4cm]{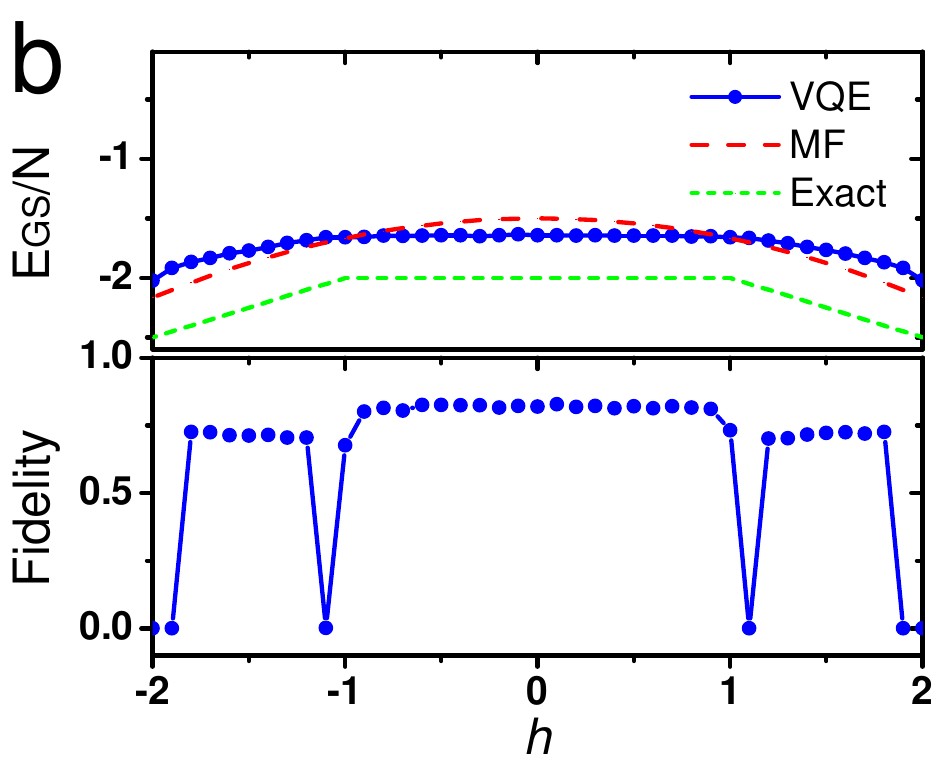}\includegraphics[width=4cm]{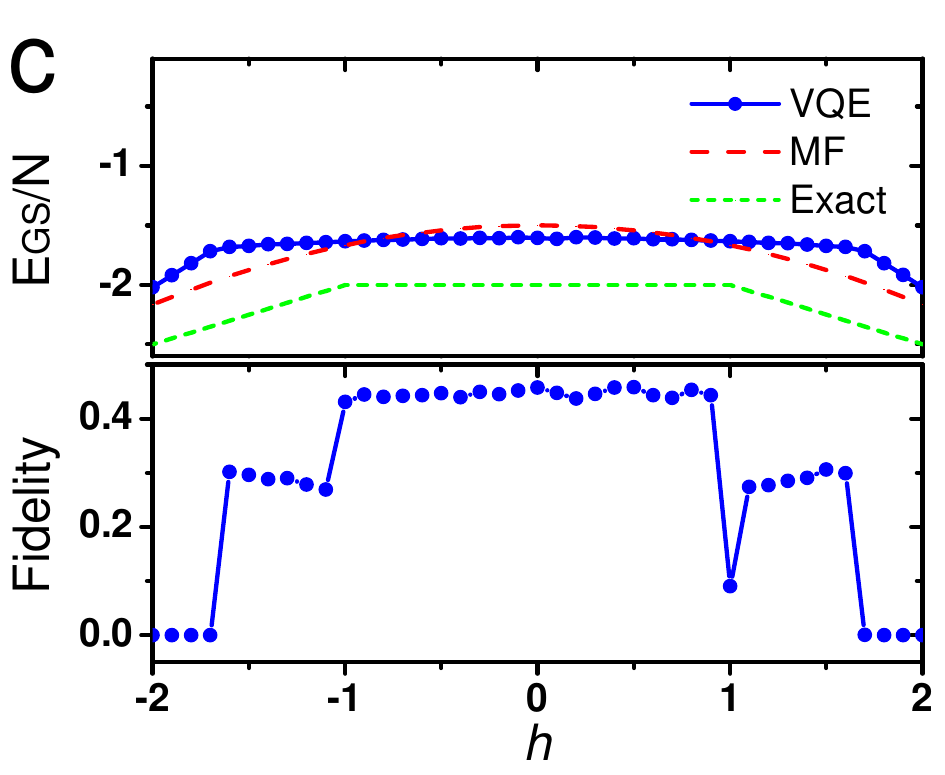}\\
\caption{VQE algorithm with CNOT ansatz. (a) quantum circuits for the linear and the full CNOT ansatzes. (b) and (c) are ground state energy per site $E_{\text{GS}}/N$ and fidelity between VQE algorithm and exact solution as a function of Zeeman field $h$ for the full and the linear ansatzes respectively. The parameters are $J=1$, $N=4$.}
\label{fig3}
\end{figure}

We find that the mean field VQE ansatz is consistent with the analytical mean-field solution, but deviates significantly from the exact solution. This is because there is some entanglement in the ground state of the model which is not captured by the mean field ansatz since the product state indicates no entanglement in the ansatz. We thus consider adding some degree of entanglement into the system by adding entanglers into the circuits as shown in Fig.~\ref{fig3}a. Here we consider adding a full CNOT entangler or a linear CNOT entangler as an entangler after the mean-field block $U_{\text{MF}}(\vec{\theta})$ has been applied. The full CNOT entangler includes a series of CNOT gates linking all the pairs of qubits while the linear one only links the nearest neighbor pairs of qubits.
\addedStart{}Note that there are no variational parameters in the CNOT entanglers; the variational parameters in the mean-field block are optimized for the CNOT ansatz.\addedEnd{}
Due to the entangler, the CNOT ansatz can no longer be expressed as a product state. We should instead directly evaluate the expectation of the energy as $\langle{H}\rangle=-J\sum_{i<j}(\langle{X_{i}X_{j}}\rangle+\langle{Y_{i}Y_{j}}\rangle)-h\sum_{i}\langle{Z_{i}}\rangle$, where the correlators of Pauli matrices are included.

The results for the full and linear CNOT ansatzes are shown in Fig.~\ref{fig3}b and \ref{fig3}c respectively. For the full CNOT ansatz, it is found that the VQE energy is still significantly different from the exact solution, implying the poor expressive power of the full CNOT ansatz for the long-range XY model. However, since the VQE energy for the ferromagnetic phase ($|h|<1$) is between the mean-field solution and exact solution, the ansatz can capture part of the entanglement in the ferromagnetic phase; while for the paramagnetic phase ($|h|>1$), the VQE energy is even worse than the mean-field solution. The fidelity is lower than 0.9 in both phases. The result for linear CNOT ansatz is shown in Fig.~\ref{fig3}c. We find that in the paramagnetic phase, the VQE energy for linear CNOT ansatz is higher than that for the full CNOT ansatz, implying that the full entangler has better expressive power than its linear counterpart. This can also be reflected in the lower fidelity ($<0.4$) for the linear ansatz. In the ferromagnetic phase, the full and linear ansatzes have similar capacity to minimize the VQE energy.

\section{VQE algorithm with CRX ansatz}

\begin{figure}[tbp]
\includegraphics[width=7cm]{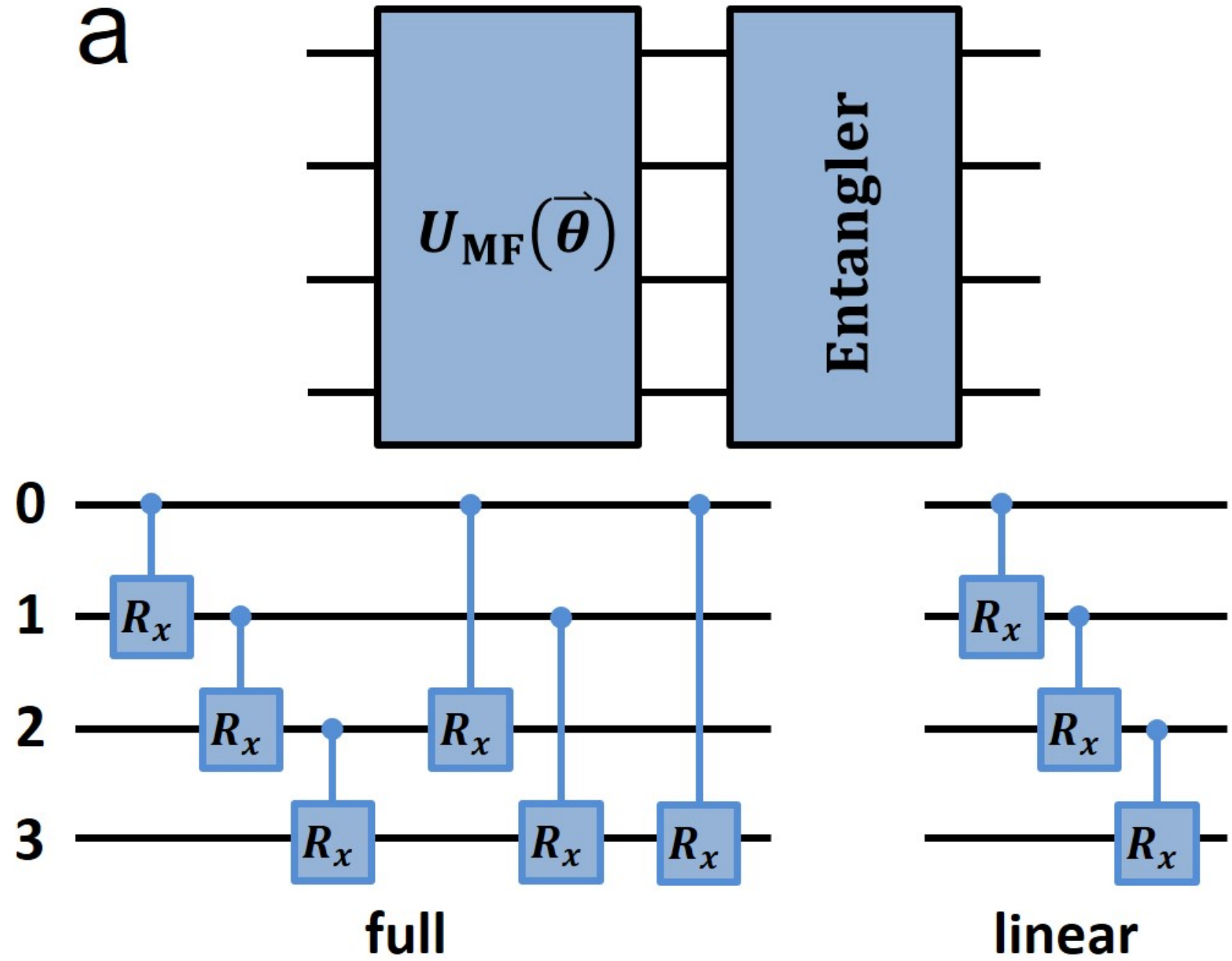}\\
\includegraphics[width=4cm]{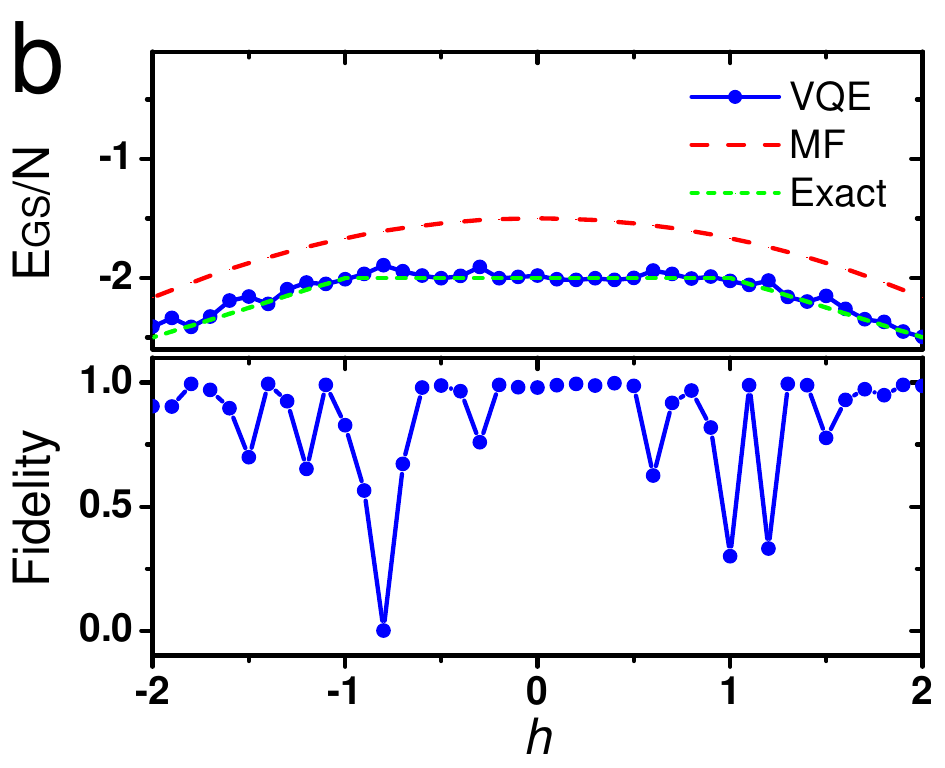}\includegraphics[width=4cm]{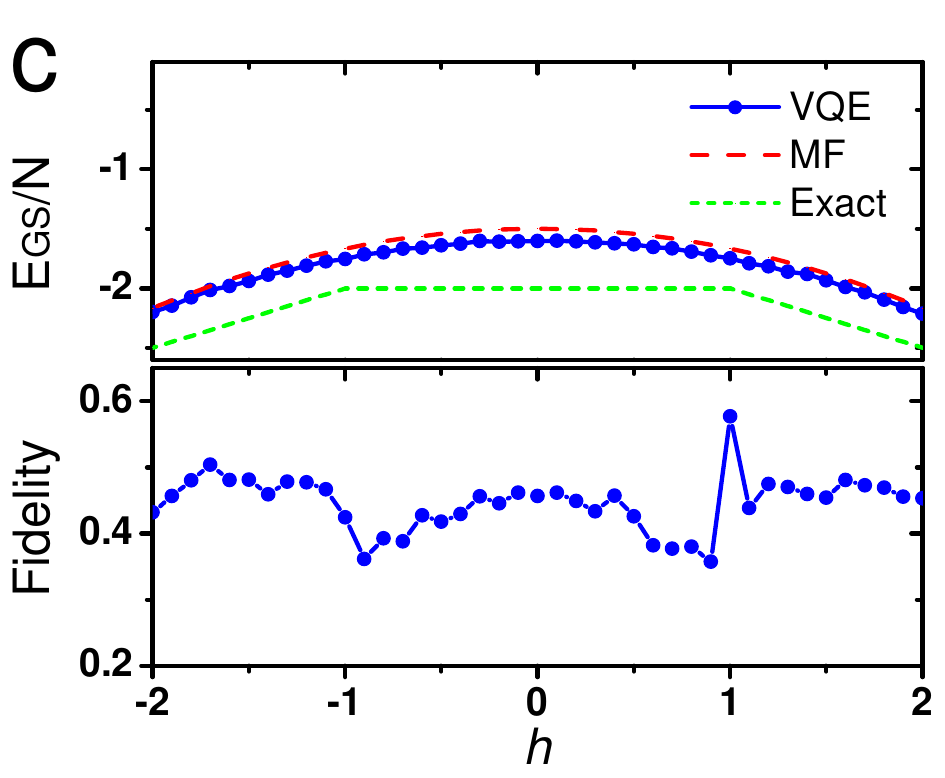}\\
\caption{VQE algorithm with CRX ansatz. a: quantum circuits for the linear and the full CRX ansatzes. b and c are ground state energy per site $E_{\text{GS}}/N$ and fidelity between VQE algorithm and exact solution as a function of Zeeman field $h$ for the full and the linear ansatzes respectively. The parameters are $J=1$, $N=4$.}
\label{fig4}
\end{figure}

We find that the CNOT ansatzes cannot approximate the ground state energy of the long-range XY model well. It is because the CNOT entanglers, having no variational parameters, have limitations in bringing entanglement into the ansatzes. To improve the ansatzes, we can introduce variational parameters into the CNOT entanglers to gradually introduce the entanglement into the ansatzes. A natural way is to generalize the CNOT gates to the controlled-rotation gates, for example, the controlled-$R_{x}$ gate, to tune the degree of entanglement more flexibly. We thus consider the full and the linear CRX entanglers as shown in Fig.~\ref{fig4}a.

In Fig.~\ref{fig4}b, we consider the full CRX ansatz and find that this ansatz significantly improves the accuracy. The VQE energy almost approaches the exact solution with some fluctuation resulting from the sampling error and the intrinsic expressive power of the ansatz. This can also be observed in the fidelity. It is found that the more the VQE energy deviates from the exact solution, the lower the fidelity. We also find that the fidelity is low near the critical point $h=1$, implying poor expressive power near the phase transition. As a comparison, we consider the linear CRX ansatz shown in Fig.~\ref{fig4}c. We find that the VQE energy of the linear ansatz has lower expressive power than the full one to describe the ground state energy of the long-range XY model. It only approaches to the mean-field solution and deviates from the exact solution. This can also be reflected in the low fidelity shown in the figure.

\section{VQE algorithm with TQR ansatz}

\begin{figure}[tbp]
\includegraphics[width=7cm]{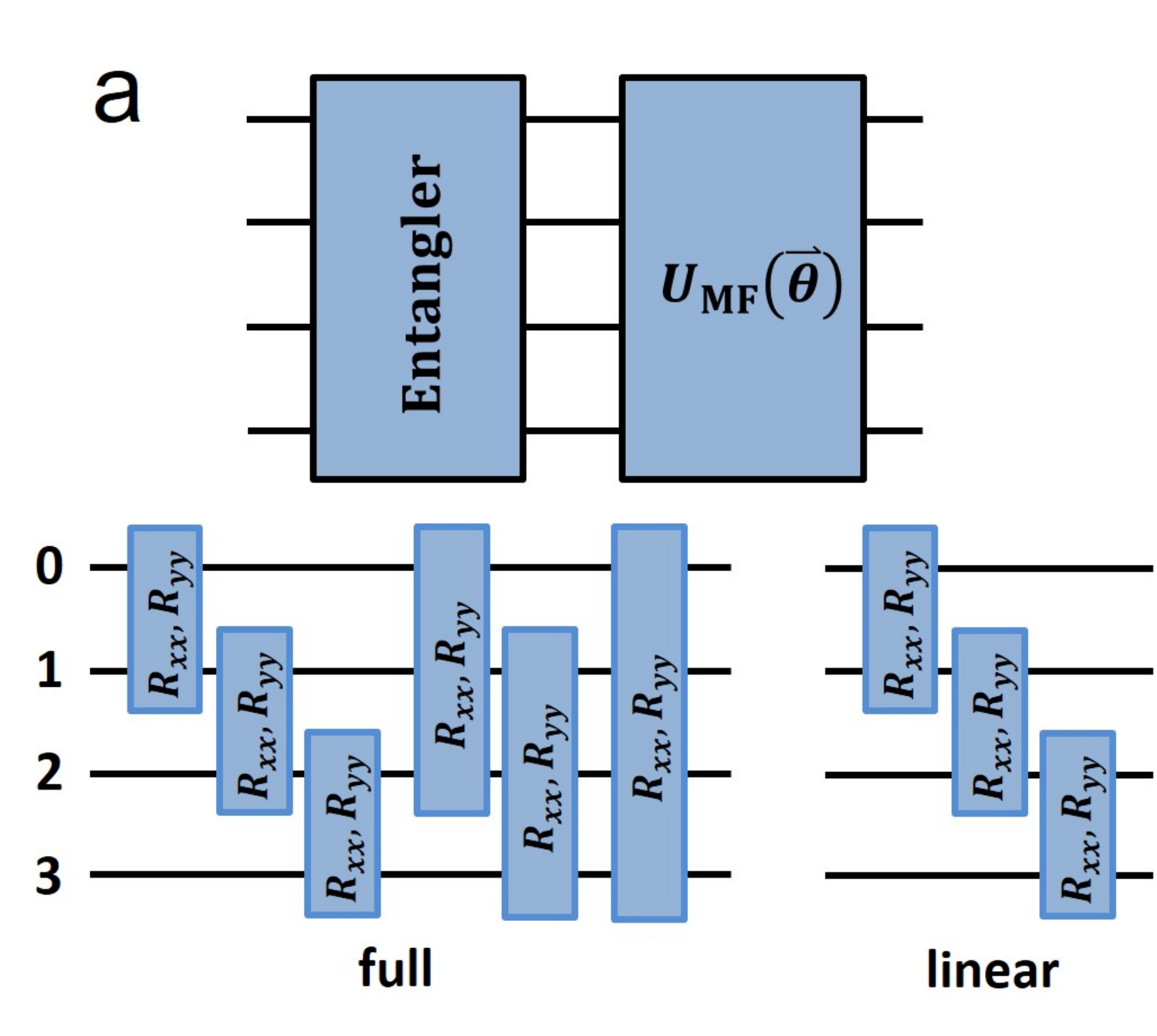}\\
\includegraphics[width=4cm]{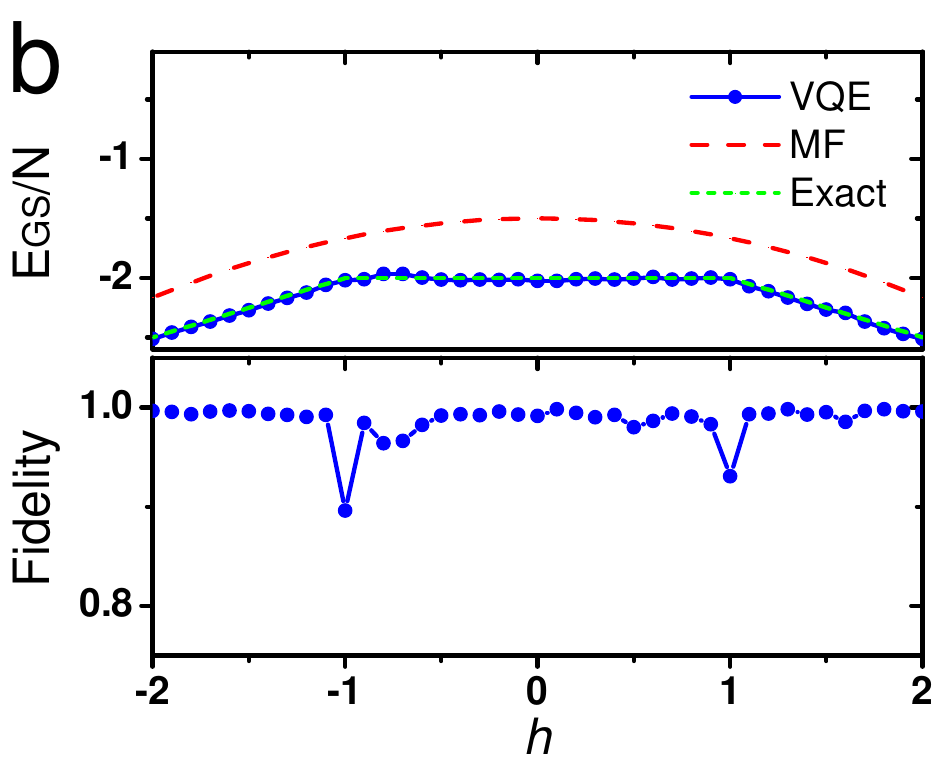}\includegraphics[width=4cm]{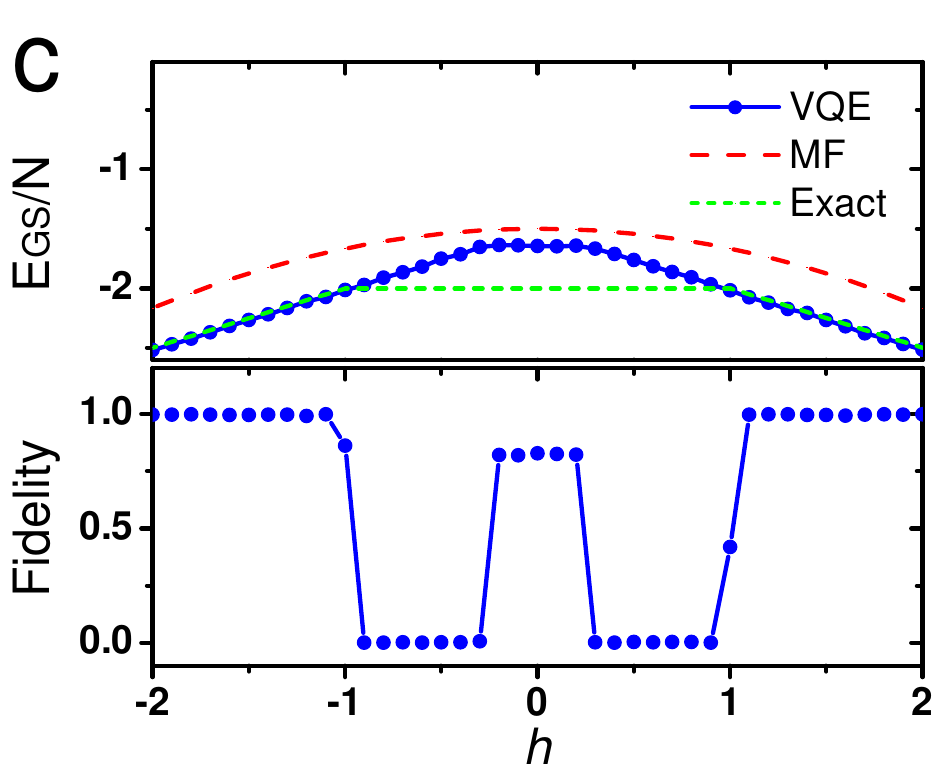}\\
\caption{VQE algorithm with the TQR ansatz. a: quantum circuits for the linear- and the full-entanglement TQR ansatzes. The box labeled ``$R_{xx},R_{yy}$'' in the circuit diagram is meant to represent the operation of performing the $R_{xx}$ and $R_{yy}$ gates in sequence. Also, note that unlike previous diagrams, the entangler is performed before the mean-field ansatz. It turns out that this performs better than the case where the order is flipped. b and c are ground state energy per site $E_{\text{GS}}/N$ and fidelity between VQE algorithm and exact solution as a function of Zeeman field $h$ for the full and the linear ansatzes respectively. The parameters are $J=1$, $N=4$.}
\label{fig5}
\end{figure}

To further improve the performance of the VQE ansatz, we propose a problem-inspired ansatz in Fig.~\ref{fig5}a based on the two-qubit rotation gates $R_{xx}(\alpha_{ij})$ and $R_{yy}(\beta_{ij})$, which are defined as $R_{xx}(\alpha_{ij})=e^{-i\alpha_{ij}{X_{i}{\otimes}X_{j}}/2}$ and $R_{yy}(\beta_{ij})=e^{-i\beta_{ij}{Y_{i}{\otimes}Y_{j}}/2}$ respectively. We consider the full and the linear TQR ansatzes for the long-range XY model.

We demonstrate the results for the full TQR ansatz in Fig.~\ref{fig5}b. It is found that this ansatz can fully capture the properties of the long-range XY model both in the ferromagnetic and the paramagnetic phases. The VQE energy for the ground state is almost the same as the exact solution; furthermore, compared with the full CRX ansatz, the fidelity is close to one except for the critical points $|h|=1$. In Fig.~\ref{fig5}c, we show the result for the linear TQR ansatz. It is found that the linear TQR ansatz is a good candidate for the paramagnetic phase of the long-range XY model which can be demonstrated by the VQE energy nearly identical to the exact solution and the fidelity close to one. However, the linear TQR ansatz is not expressive enough to represent a state in the ferromagnetic phase. The VQE energy deviates from the plateau in the exact solution in the ferromagnetic phase. We observe that a new plateau is developed with lower fidelity in the ferromagnetic phase. Strikingly, apart from the plateau in the VQE energy, the fidelity even drops to zero. However, the VQE energy is in between the mean-field solution and the exact solution, implying that some degree of entanglement is captured by the ansatz. Comparing with the linear CRX ansatz shown in Fig.~\ref{fig4}d, the linear TQR ansatz performs better than the linear CRX ansatz.

\addedStart{}
\section{Gate order dependence of different VQE ansatzes}
\label{sec:gate_order_dependence}

Note that the two-qubit gates in the entanglers of the three VQE ansatzes (CNOT, CRX and TQR) do not commute with each other, and so a natural question is to study whether the gate order in the entangler affects the VQE results. For the four-qubit VQE ansatzes with full entanglers, there are $6!\times2^6=46080$ ways to arrange the gates in the circuit. We do not exhaust all the possibilities here but randomly choose 4 (10) realizations for full CNOT (CRX, TQR) ansatzes to investigate the dependence on the gate order.

\begin{figure}[tbp]
\includegraphics[width=8cm]{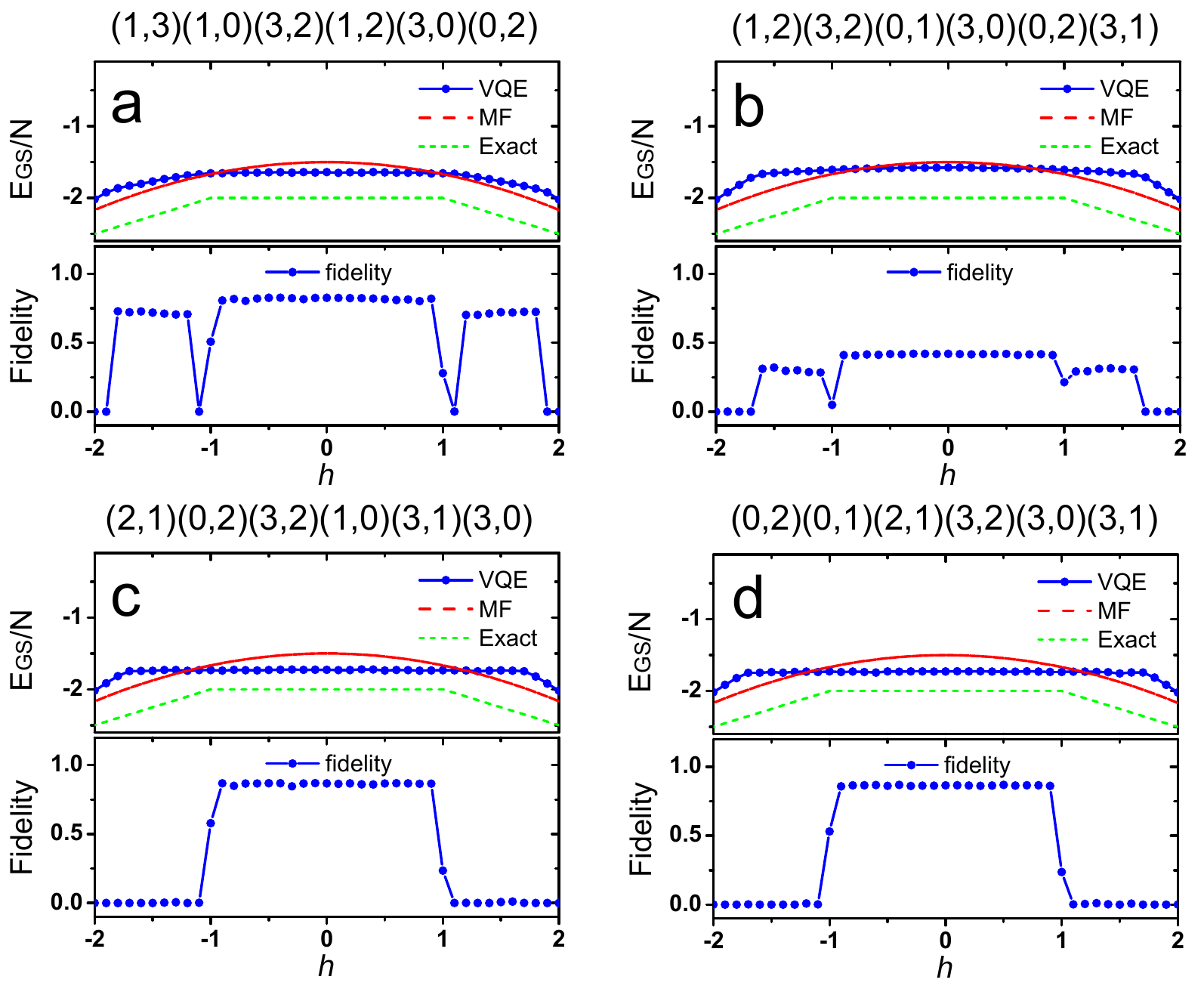}\\
\caption{Gate order dependence of the full CNOT ansatz. Subfigures a--d are ground state energy
per site $E_{\text{GS}}/N$ and fidelity between VQE algorithm and exact solution as a function of Zeeman field $h$ for the four random realizations of the full CNOT ansatz. The parameters are $J=1$, $N=4$.}
\label{figs1}
\end{figure}
The gate order dependence for the full CNOT ansatz is shown in Fig.~\ref{figs1}. For each realization, we obtain the VQE ground state energy and fidelity. The gate order in the realization is shown in the title of each figure. For example, $(1,3)(1,0)(3,2)(1,2)(3,0)(0,2)$ in Fig.~\ref{figs1}a means that we consecutively append CNOT gates to the
circuit in the order specified. The notation $(i,j)$ means that we apply the CNOT gate to the qubit pair $(i,j)$, with $i$ being the control qubit and $j$ being the target qubit. It is shown that for the full CNOT ansatz, the VQE energy and fidelity are indeed dependent on the gate order and are always far from the exact energy. This is due to the insufficient entanglement inside the full CNOT ansatz, which we will illustrate in detail in Sec.~\ref{expressibility}. 

\begin{figure*}[tbp]
\includegraphics[width=18cm]{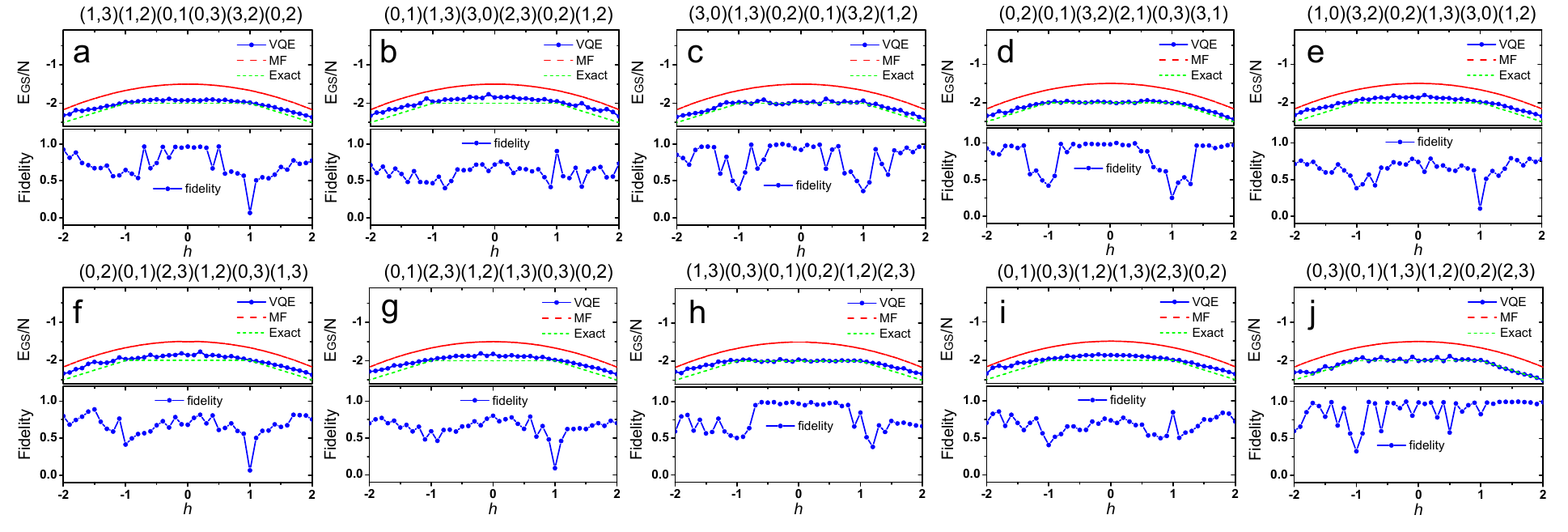}\\
\caption{Gate order dependence of the full CRX ansatz. a-j are ground state energy
per site $E_{\text{GS}}/N$ and fidelity between VQE algorithm and exact solution as a function of Zeeman field $h$ for the ten random realizations of the full CRX ansatz. The parameters are $J=1$, $N=4$.}
\label{figs2}
\end{figure*}
The gate order dependence for the full CRX ansatz is shown in Fig.~\ref{figs2}. Since the full CRX ansatz contains more variational parameters than the full CNOT ansatz, we sample ten realizations for the full CRX ansatz. Compared to the full CNOT ansatz, it is found that while some of the realizations can well approximate the exact ground energy, from the view of the fidelity, no realizations can well describe the ground state wavefunction of the model.

\begin{figure*}[tbp]
\includegraphics[width=18cm]{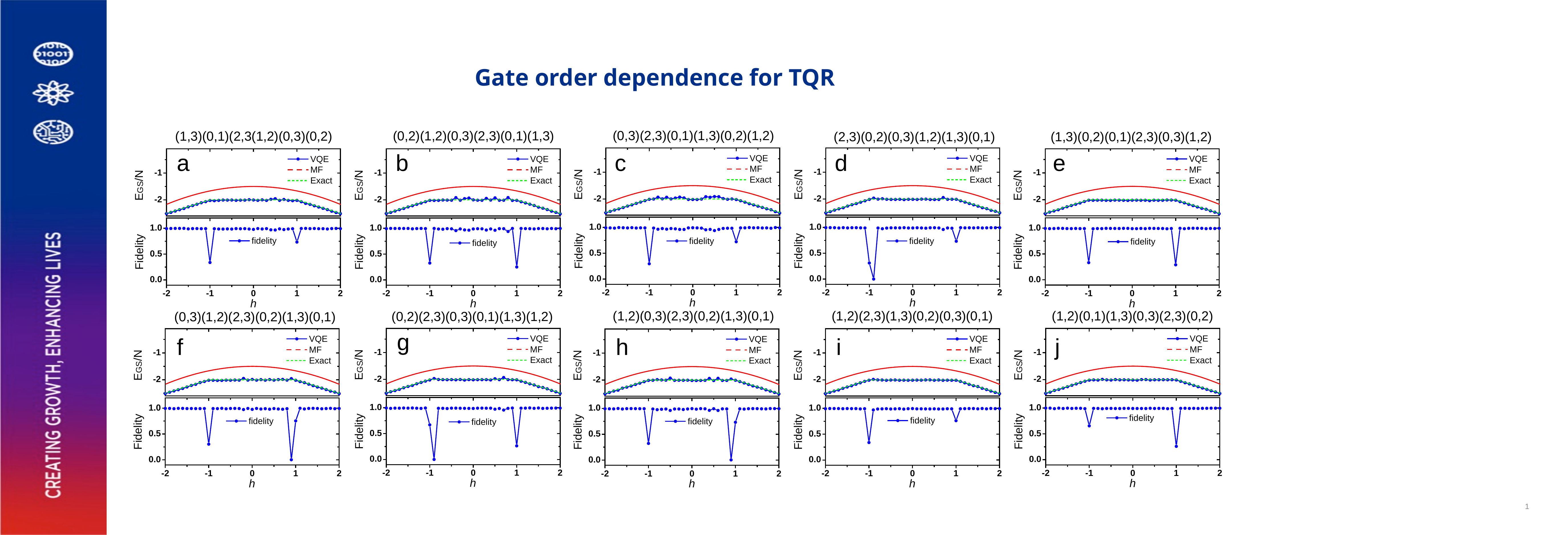}\\
\caption{Gate order dependence of the full TQR ansatz. a-j are ground state energy
per site $E_{\text{GS}}/N$ and fidelity between VQE algorithm and exact solution as a function of Zeeman field $h$ for the ten random realizations of the full TQR ansatz. The parameters are $J=1$, $N=4$.}
\label{figs3}
\end{figure*}
The gate order dependence for the full TQR ansatz is shown in Fig.~\ref{figs3}. We also give ten realizations for the full TQR ansatz. We find that among the CNOT, CRX and TQR ansatzes, only TQR ansatz can satisfactorily describe both the ground state energy and the wavefunction of the model. Furthermore, it is shown that except for the points near the phase boundary $|h|=1$, the gate order has no effect on the TQR ansatz, unlike the case for the CNOT and CRX ansatzes.

\section{Multi-layer setup for different VQE ansatzes}
\label{sec:multi-layer_setup}

\begin{figure}[tbp]
\includegraphics[width=8cm]{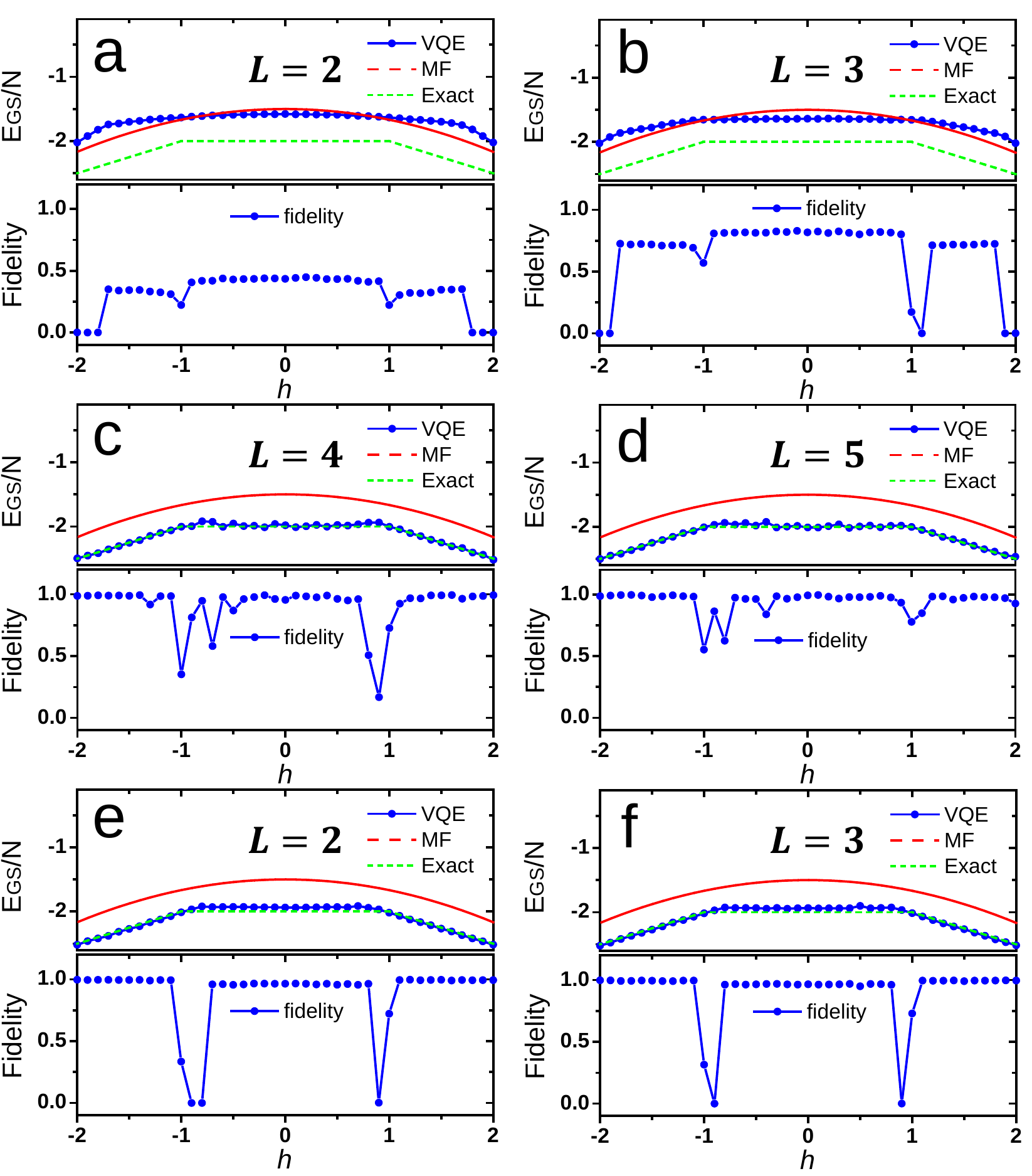}\\
\caption{Multi-layer setup for the three full entangler ansatzes. a,b: layer dependence ($L=2,3$) for full CNOT ansatz. c,d: layer dependence ($L=4,5$) for full CRX ansatz. e,f: layer dependence ($L=2,3$) for full TQR ansatz. The parameters are $J=1$, $N=4$.}
\label{figs4}
\end{figure}
In this section, we will study the multi-layer setup for all the three ansatzes. To this end, we repeat the entangler in each of the three full ansatzes (CNOT, CRX and TQR) $L$ times with independent parameters for each layer and study the VQE energy and fidelity with an increasing number of layers (entanglers).

In Fig.~\ref{figs4}, we present the layer dependence for the three full entangler ansatzes. In Figs.~\ref{figs4}a and \ref{figs4}b, we find that the full CNOT ansatz cannot approach the exact ground state energy and wavefunction in spite of the increase in the number of layers. This will be further explained from the point of view of the entanglement entropy in Sec.~\ref{expressibility}. 

In Figs.~\ref{figs4}c and \ref{figs4}d, we find that the full CRX ansatz can approximate the exact ground state energy and wavefunction when the number of layers $L\ge4$. However, there are some fluctuations for the VQE wavefunction near the phase boundary $|h|=1$, which can be mitigated by increasing the number of layers. Again, we find that the best multi-layer ansatz for the long-range XY model is the full TQR ansatz. Even for the single-layer setup in Fig.~\ref{fig5}, the TQR ansatz can well describe the ground state energy and wavefunction. Furthermore, from Fig.~\ref{figs4}e and \ref{figs4}f, the stability improves with the increase in the number of layers.
\addedEnd{}

\section{Expressive power of different VQE ansatzes}
\label{expressibility}

In the above sections, we find that there are significant differences among the VQE ansatzes in representing the ground state of the long-range XY model. Here we will characterize the expressive power of different ansatzes from the point of view of the entanglement entropy \cite{PhysRevX.7.031016,https://doi.org/10.1002/qute.201900070}. Consider an $N$-qubit chain with sites labeled $0,1,2,\ldots,N-1$. We assume open boundary conditions and label the bonds of the chain by $x$ ($0\le{x}\le{N}$). Here $x=0$ and $x=N$ are two virtual bonds to the left of qubit $0$ and to the right of qubit $N-1$ respectively. Since the quantum circuits consist of only unitary gates, the full density matrix $\rho=|\Psi\rangle\!\langle\Psi|$ is a pure state. We now consider the entanglement entropy across a single cut of the bond at position $x$. We first define the reduced density matrix $\rho_{x}$ by chopping the chain into two parts at $x$ and tracing out the right-hand side. The entanglement entropy is defined as the von Neumann entropy for a cut at $x$, which is given by $S(x)=-\text{Tr}\rho_{x}\log_{2}\rho_{x}$. Note that there are boundary conditions on the entropy: $S(0)=S(N)=0$.

\begin{figure}[tbp]
\includegraphics[width=4.2cm]{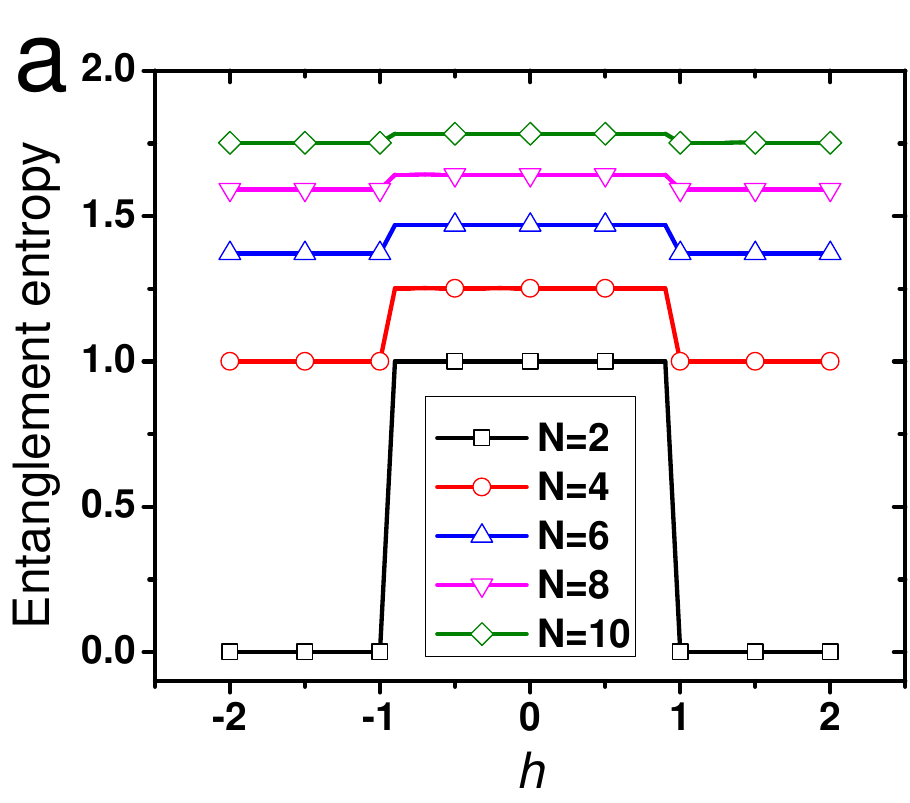}\includegraphics[width=4.2cm]{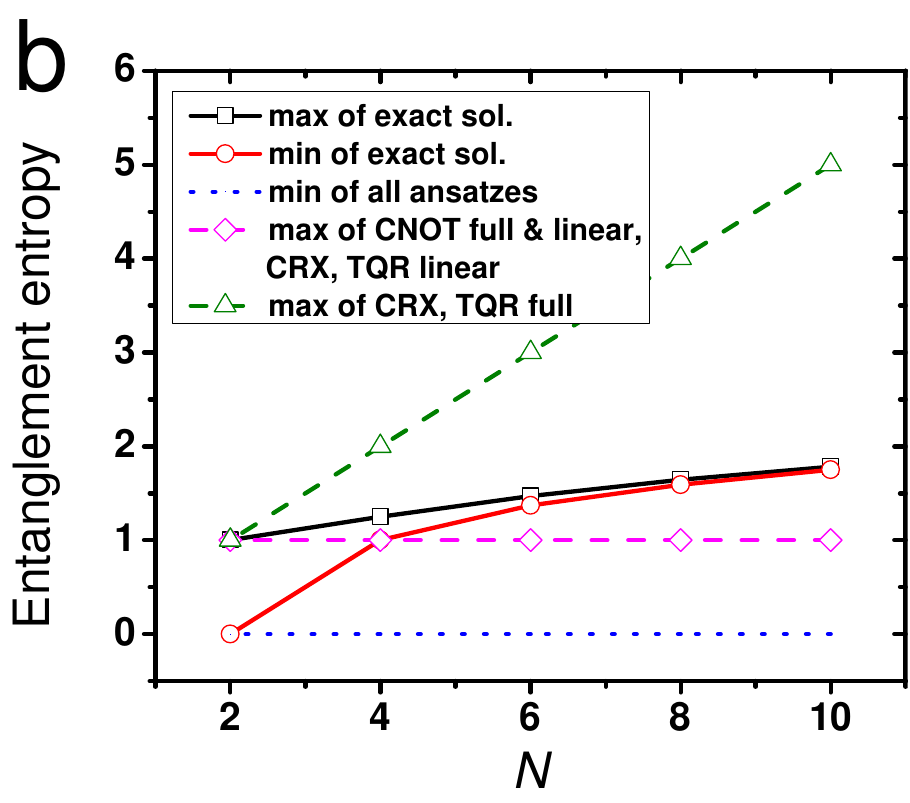}\\
\includegraphics[width=4.3cm]{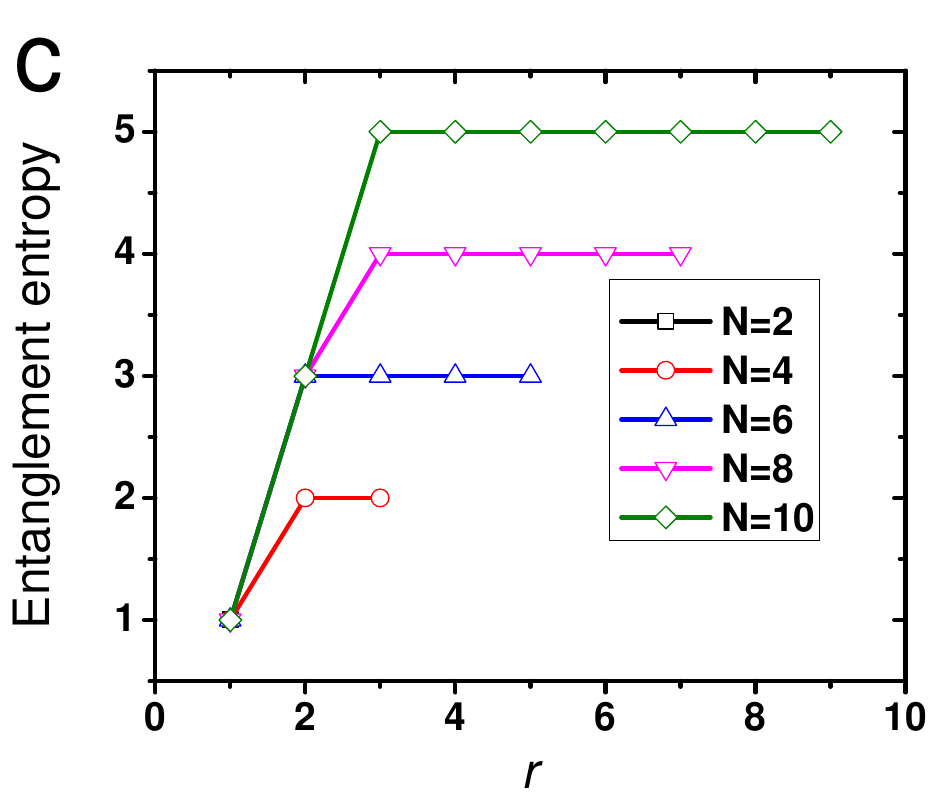}\includegraphics[width=4.4cm]{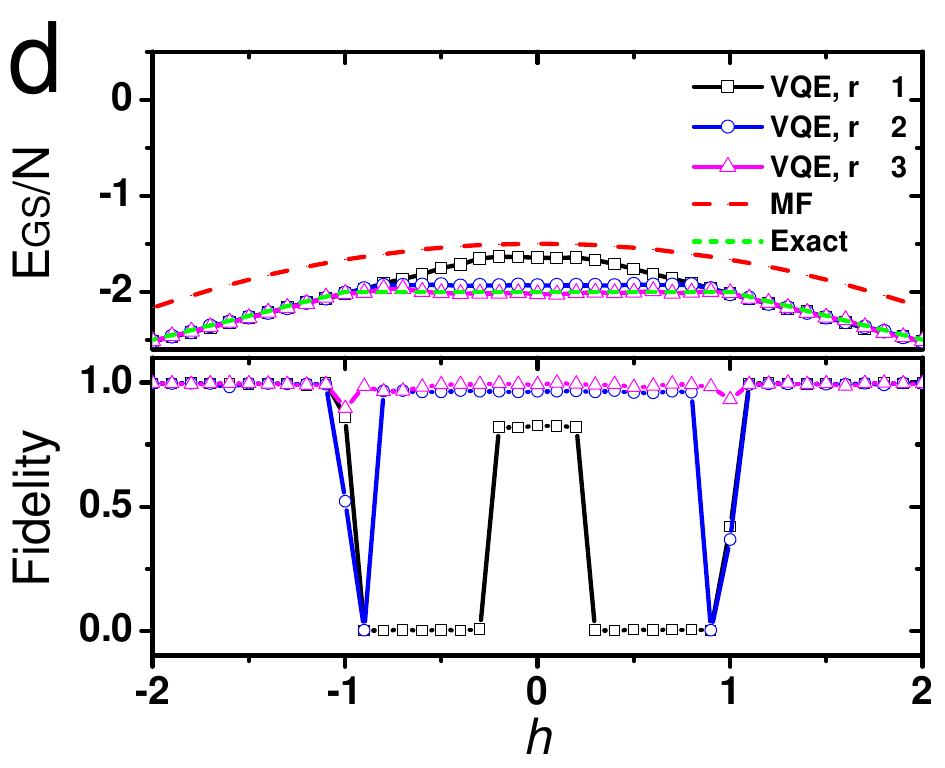}\\
    \caption{(a) Half chain entanglement entropy as a function of Zeeman field $h$ for the exact solution. The entropy has centain values to the left and right of the critical points $|h|=1$. (b) Half chain entanglement entropy as a function of number of qubits $N$ (even) for the exact solution and different VQE ansatzes. The black solid ($\square$) and red solid ($\bigcirc$) lines are ferromagnetic and paramagnetic entropies for the exact solution of the model. The blue dotted line is the minimum entropy for all VQE ansatzes. The pink dashed ($\Diamond$) line is the maximum entropy for the linear and full CNOT entanglers, the linear CRX and TQR entanglers. The olive triangle ($\bigtriangleup$) line is the maximum entropy for the full CRX and TQR entanglers. (c) Maximum half chain entanglement entropy as a function of neighborhood size $r$ for different number of qubits $N$. Note that for $N=2$, the entropy is equal to 1 only when $r=1$. (d) The ground state energy per site $E_{\text{GS}}/N$ and fidelity between VQE algorithm and exact solution as a function of Zeeman field $h$ for different neighborhood $r$. The parameters are $J=1$, $N=4$.}
\label{fig6}
\end{figure}
\begin{figure}[tbp]
\includegraphics[width=4.2cm]{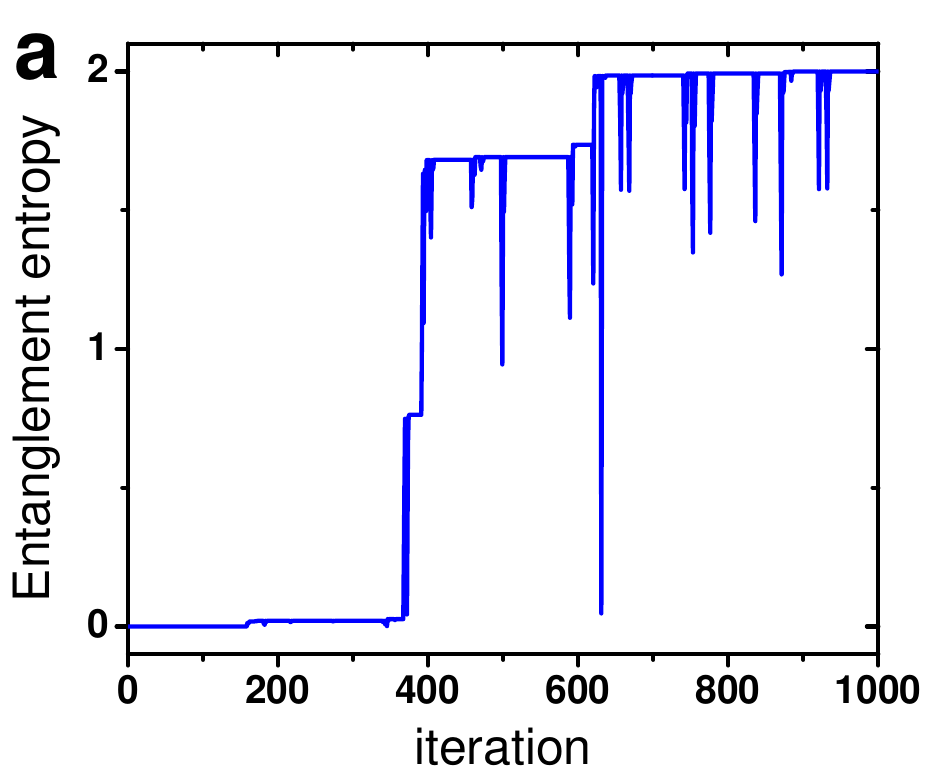}\includegraphics[width=4.1cm]{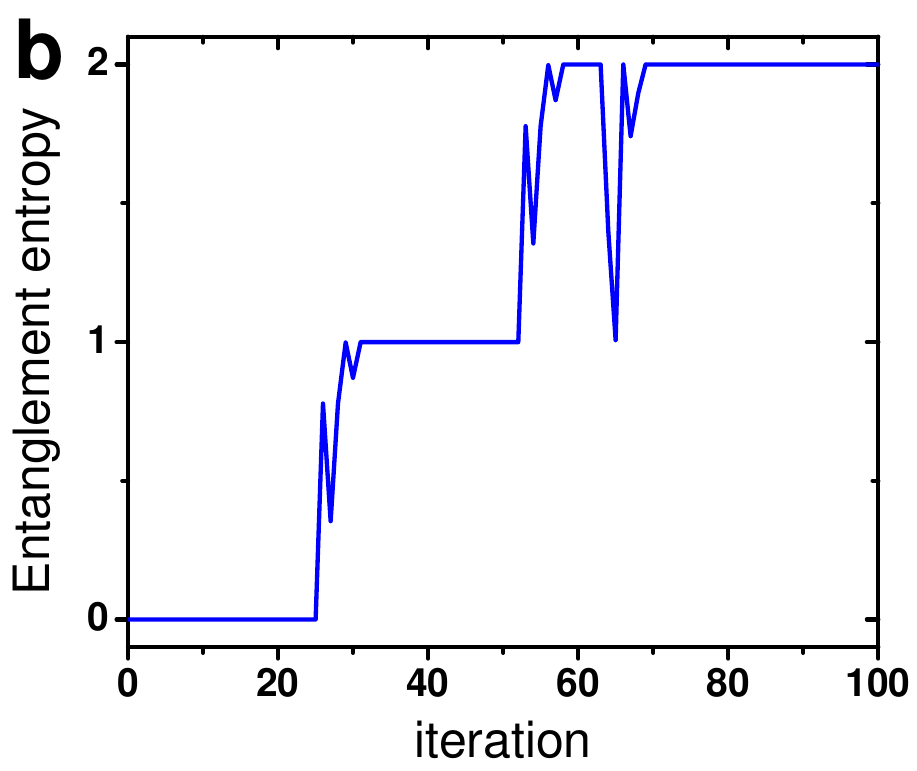}\\
\includegraphics[width=4.1cm]{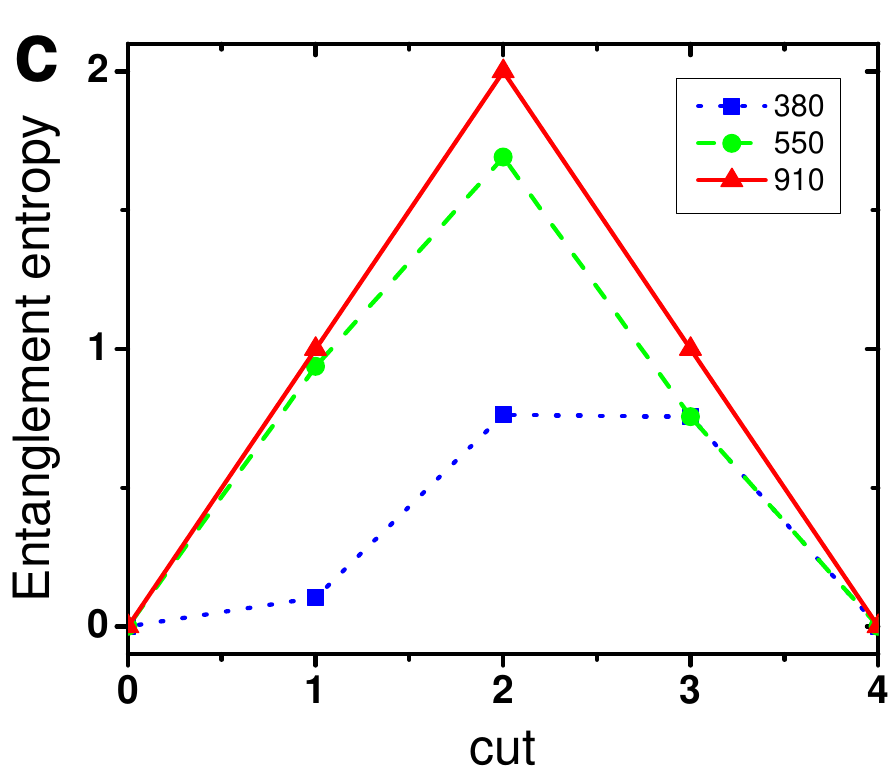}\includegraphics[width=4.1cm]{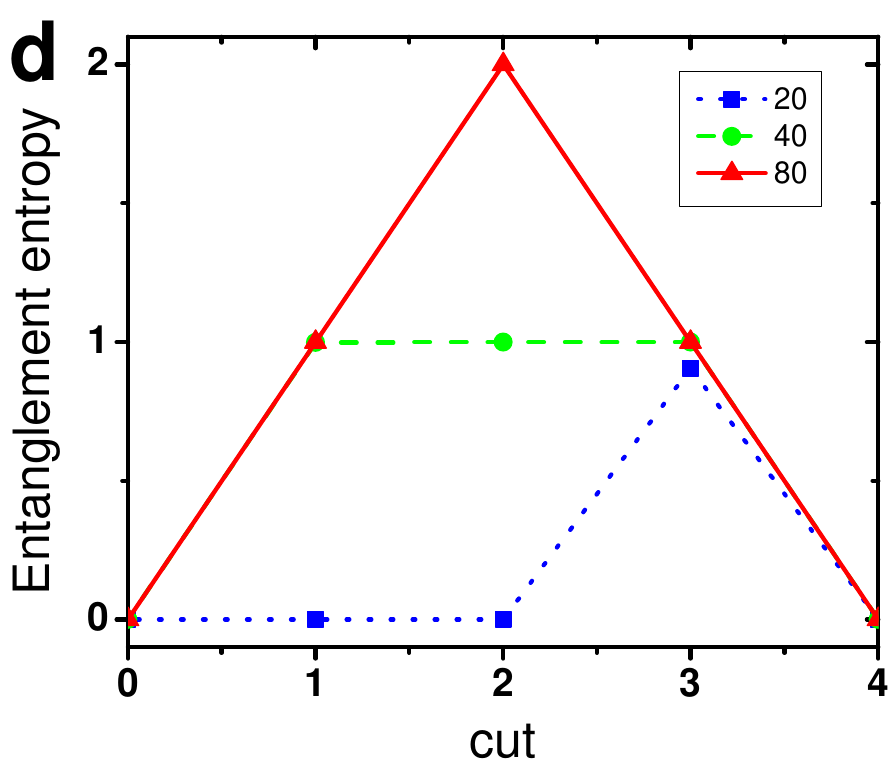}\\
\caption{The growth of entanglement entropy for a chain with length $N=4$. (a, b) Half chain entanglement entropy as a function of iteration $n$ for the full CRX and TQR entanglers.
(c, d) Entanglement entropy as a function of cut $x$, for several successive iterations ($n=380, 550, 910$ for CRX entangler and $n=20, 40, 80$ for TQR entangler). This shows that the state evolves from a product state to a maximally-entangled one.}
\label{fig7}
\end{figure}
For even $N$, we first study the half chain entropy $S(N/2)$ as a function of the Zeeman field $h$ for the long-range XY model as shown in Fig.~\ref{fig6}a. It is found that the entropy remains a constant within the same quantum phase while has a jump at the phase boundary. The entropy also increases with the number of qubits $N$ increases. We summarize the paramagnetic  (low value) and ferromagnetic (high value) entropies for different $N$ in Fig.~\ref{fig6}a and shown in the black solid ($\square$) and red solid ($\bigcirc$) lines in Fig.~\ref{fig6}b.

We then study the half chain entropy for the VQE ansatzes as shown in Fig.~\ref{fig6}b. It is easy to check that the minimum entropy is zero for all ansatzes since the variational parameters in the ansatzes can be tuned to make the resulting state become separable. The maximum entropy for the linear and full CNOT entanglers, and the linear CRX and TQR entanglers is $1$ while $N/2$ for the full CRX and TQR entanglers which is the largest possible entropy for an $N$-qubit chain. Therefore, from the entanglement entropy point of view, it is clear that only the full CRX and TQR entanglers have enough expressive power to represent the ground state of the long-range XY model. This has been shown in the previous sections.

We find that the convergence accuracy of the ground state energy can be improved by gradually introducing the two-qubit rotation gates $R_{xx}(\alpha_{ij})$ and $R_{yy}(\beta_{ij})$ into the circuits. We first define the distance of a pair of qubits by $r_{ij}=|i-j|$, where $i,j$ are the index of qubits. Then we create the entanglers by including the two-qubit rotation gates $R_{xx}(\alpha_{ij})$ and $R_{yy}(\beta_{ij})$ with the distance $r_{ij}\le{r}$. In Fig.~\ref{fig6}c, it is found that for a circuit with number of qubits $N$, the half-chain entropy can reach to the maximal entropy $N/2$ before the full entangler is applied. Therefore, it is possible to obtain an acceptable solution without taking into account the full entangler and this has the advantage of faster simulation time. In Fig.~\ref{fig6}d, we plot the ground state energy and fidelity for $N=4$ and for different neighborhood $r$, and find that when $r$ goes from $1$ (linear) to $3$ (full), the accuracies of ground state energy and fidelity are gradually improved. An acceptable solution for the long-range XY model can be achieved when $r=2$.

We will further interpret the expressive powers of different VQE ansatzes, especially the full CRX and TQR ansatzes, from a dynamical point of view. In Fig.~\ref{fig7}a and \ref{fig7}b, we study the entanglement entropy growth (half chain entropy) as a function of the number of iteration for the full CRX and TQR ansatzes. The optimization object is to maximize the half chain entropy $S(N/2)$ and is done by the \texttt{Powell} method in the \texttt{minimize} routine of the \texttt{SciPy} package. It is found that the full TQR ansatz converges about one magnitude faster than the full CRX ansatz. In Fig.~\ref{fig7}c and \ref{fig7}d, we give the growth of the entanglement entropy as a function of the cut $x$, for several successive iterations. It is found that the maximal entropy profile, $S_\text{max}(x)=\big|\left|x-N/2\right|-N/2\big|$ can be reached for both the full CRX and full TQR ansatzes; however, the full TQR ansatz approaches faster to the maximally-entangled state than the full CRX ansatz. In addition, we find that the transient entropy profiles are asymmetric. This is because the transient step is for a particular realization of the quantum circuit with circuit parameters randomly distributed in the parameter space. However, the profile will eventually go to the maximally-entangled state with a symmetric profile.

\section{Conclusion}

In this paper, we studied the ground state of the long-range XY model by utilizing the VQE algorithm. Among different VQE ansatzes, i.e., mean-field, linear and full CNOT, CRX, and TQR ansatzes, we found that the full CRX and TQR ansatzes can represent the ground state energy of the long-range XY model well. In contrast, only the full TQR ansatz can represent the ground state wavefunction with a fidelity close to one.\addedStart{} 
The TQR ansatz can reduce the circuit depth by increasing the number of variational parameters introduced in each layer. This is practical for NISQ devices 
where the noise grows rapidly with circuit depth and affects the fidelity of the prepared quantum state.
\addedEnd{}We also found that an acceptable solution can be achieved using restricted-entanglement ansatzes where entangling gates are applied only between qubits that are a fixed distance from each other. Using these restricted-entanglement ansatzes instead of full-entanglement ansatzes allows for faster simulation. Finally, we discuss the performance of VQE ansatzes from the point of view of the entanglement entropy and find that the full TQR ansatz is more powerful in representing the ground state energy and wavefunction of the long-range XY model than the full CRX ansatz.

\section*{Acknowledgments}

We would like to thank Mirko Consiglio and Tony J.\ G.\ Apollaro for useful discussions. 
The IHPC A*STAR Team acknowledges support from the A*STAR Career Development Award (C210112010), A*STAR (\#21709), National Research Foundation Singapore (NRF2021-QEP2-02-P01, QEP-SF1, NRF2021-QEP2-02-P03) and A*STAR Computational Resource Centre through the use of its high performance computing facilities. Wu gratefully acknowledges the Start-Up Research Grant from Singapore University of Technology and Design via Grant No.~SRG SMT 2021 169, and National Research Foundation Singapore via Grant No.~NRF2021-QEP2-03-P09 and NRF-CRP26-2021-0004. We acknowledge the use of IBM Quantum services for this work. The views expressed are those of the authors, and do not reflect the official policy or position of IBM or the IBM Quantum team.

%\bibliography{biblio}


\end{document}